\documentclass[conference]{IEEEtran}
\usepackage[utf8]{inputenc}
\usepackage[utf8]{inputenc}
\usepackage[english]{babel}
\usepackage[T1]{fontenc}
\usepackage{amsmath}
\usepackage{amsfonts}
\usepackage{amssymb}
\usepackage{url}
\usepackage[official]{eurosym}
\usepackage{pgfplots}
\pgfplotsset{width=7cm}
\usetikzlibrary{patterns}
\usepackage{tikz-cd}
\usetikzlibrary{shapes,arrows}
\usetikzlibrary{calc,fit,trees,positioning,arrows,chains,shapes.geometric,shapes}
\usetikzlibrary{shapes,arrows}
\author{\IEEEauthorblockN{Ricardo Morla}
\IEEEauthorblockA{ricardo.morla@fe.up.pt\\ INESC TEC, Faculty of Engineering, University of Porto\\Porto, Portugal
}}
\title{Effect of Pipelining and Multiplexing\\in Estimating HTTP/2.0 Web Object Sizes}
\date{May-July 2016}

\RequirePackage{filecontents}

\begin{document}

\maketitle
\thispagestyle{plain}
\pagestyle{plain}

\begin{abstract}
HTTP response size is a well-known side channel attack. With the deployment of HTTP/2.0, response size estimation attacks are generally dismissed with the argument that pipelining and response multiplexing prevent eavesdroppers from finding out response sizes. Yet the impact that pipelining and response multiplexing actually have in estimating HTTP response sizes has not been adequately investigated. In this paper we set out to help understand the effect of pipelining and response multiplexing in estimating the size of web objects on the Internet. We conduct an experiment that collects HTTP response sizes and TLS record sizes from 10k popular web sites. We gather evidence on and discuss reasons for the limited amount of pipelining and response multiplexing used on the Internet today: only 29\% of the HTTP2 web objects we observe are pipelined and only 5\% multiplexed. We also provide worst case results under different attack assumptions and show how effective a simple model for estimating response sizes from TLS record sizes can be. Our conclusion is that pipelining and especially response multiplexing can yield, as expected, a perceivable increase in relative object size estimation error yet the limited extent of multiplexing observed on the Internet today and the relative simplicity of attacks to the current pipelining mechanisms hinder their ability to help prevent web object size estimation.
\end{abstract}

\section{Introduction}

HTTP with TLS encryption prevents attacks that inspect HTTP payload and signaling. HTTP response size analysis is a well-known side-channel attack~\cite{hintz_fingerprinting_2002} that overcomes encrypted payload inspection by using eavesdropped sizes of web objects to identify web applications.  Up to HTTP/1.1, the web client typically waits for the response to the current HTTP request before issuing the next request. This makes it straightforward to find the size of HTTP responses by tapping into the TCP/IP connection and filtering data by client-to-server and server-to-client directions. With the deployment of HTTP/2.0~\cite{varvello_is_2016} with its pipelining and multiplexing mechanisms, most authors assume HTTP response size analysis attacks can be prevented. With request pipelining, clients no longer need to wait for the response to the current HTTP request to issue the next request. With response multiplexing, servers no longer need to wait for the end of the current response to start sending the next response. Distinguishing web object sizes by eavesdropping TLS record sizes should thus be unfeasible, or at least harder than without HTTP/2.0.

Yet the extent of the effect of pipelining and multiplexing on estimating HTTP response sizes on the Web has not been adequately investigated. The fact that these mechanisms exist and are deployed does not mean that they are used and that they have an effect in response size estimation. Web content from a web page is often not pulled from the server at once and, if it is, HTTP signaling information may leak through TLS to help the attacker. This means that the privacy of regular web site users who will not have any particular reason for using anonymity tools like Tor~\cite{he_novel_2014} may be more at risk than what is believed and that the transition to HTTP/2.0 alone may not fully prevent this risk. This is especially relevant for the growing amount of traffic that goes through proxies and content delivery networks that share IP addresses between applications and for which a simple IP database lookup would not suffice for identifying web sites and applications.

In this paper we collect network traffic that our browser generates when opening a web page from a list of popular web sites, which may or not be using the HTTP/2.0 protocol. We analyze a set of empirical statistical distributions obtained from this captured traffic and report on the details of how the different concepts of HTTP/2.0 are used and the impact this has on the ability of an attacker to estimate web object sizes without having to break the TLS encryption. We provide definitions for HTTP/2.0 concepts and for pipelining and multiplexing in section \ref{sec:definitions}, which can be useful for better understanding the rest of the paper. In section \ref{sec:observing} we provide examples of pipelining and multiplexing for a small set of web sites and describe the evolution of pipelining and multiplexing for that set of web sites throughout a month. Then we describe details of the traffic capture methodology in section \ref{sec:methodology} and characterize HTTP/2.0 traffic in section \ref{sec:traffic-char}, starting with TCP streams and HTTP/2.0 web objects and drilling down to HTTP/2.0 frames and how they're mapped into TLS records. Evidence for the limited extent of pipelining and multiplexing is presented in section \ref{sec-extent}, followed by a discussion on two potential reasons. In particular we explore the relation between pipelining and the number of web objects per TCP stream and between multiplexing and the number of HTTP/2.0 frame segments per web object. We then consider the perspective of an attacker and define three increasingly stronger attack assumptions in section \ref{sec-attack-assumptions}. We estimate web object sizes under each assumption and quantify the worst case error of an attack that does no better than meeting the assumption. Finally, we present an example of an actual attack in section \ref{sec-example-attack} where we explore TLS record size patterns for a set of server IP addresses and their relation with HTTP/2.0 signaling. We discuss related work in section \ref{sec:rw} and present our conclusions and future work in section \ref{sec:conclusions}.

\section{Definitions}
\label{sec:definitions}

\begin{figure}[h!]
\begin{center}
\includegraphics[width=0.5 \textwidth]{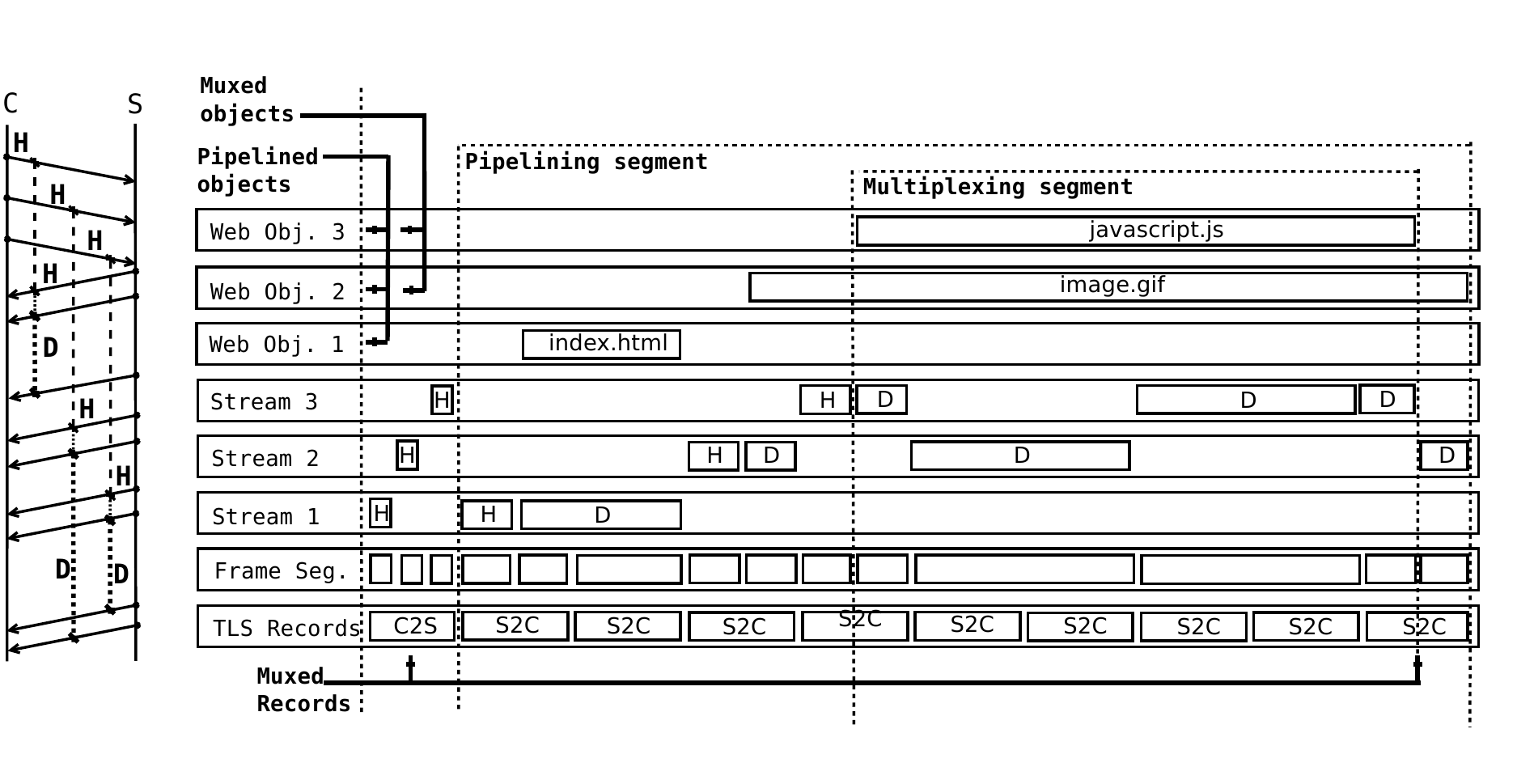} 
\end{center}
\caption{HTTP/2.0 toy example for three web objects. Left: request and response sequence diagram for the three objects. Right: TLS records and HTTP/2.0 frame segments, HTTP/2.0 streams and frames for the three web objects. C2S: client to server records, S2C: server to client records, H: header frames, D: data frames.}
\label{http2diagram}
\end{figure}

Figure \ref{http2diagram} illustrates the request-response sequence for an HTTP/2.0 toy example with three web objects. We can observe that the three web objects are pipelined since the request header for the second object is sent by the client before the response data for the first object is fully received. The same situation happens with the second and third object. We can also observe that the second and third objects are multiplexed because response data for the third web object is sent before all response data for the second object is sent by the server. This figure also illustrates mapping of web objects to frames and TLS records.

The rest of this section provides definitions related to HTTP/2.0, pipelining, and multiplexing. We also include the definition of a set of indicators related to pipelining and multiplexing that we believe are useful for observing privacy on the Internet.

\subsection{HTTP/2.0}

\subsubsection{HTTP/2.0 web object} typically HTML, script, image, or other web resource response data sent in an HTTP/2.0 stream by the server. 

\subsubsection{HTTP/2.0 stream} defines request, response, header, data bytes for a web object and also supports HTTP/2.0 signaling messages. 

\subsubsection{HTTP/2.0 frame} encapsulates request, response, header, and data as well as signalling for an HTTP/2.0 stream. Two or more HTTP/2.0 frames may be needed to send request or response bytes. 

\subsubsection{HTTP/2.0 frame segment} supports splitting frames and packaging into encrypted TLS application records. One frame segment contains an HTTP/2.0 frame or part of it. One or more HTTP/2.0 frame segments are encapsulated in TLS application records, which are then multiplexed on the TCP stream by the server and the client.

\subsection{Pipelining and Multiplexing}

\subsubsection{Pending requests and Active HTTP/2.0 streams}

Pending requests are HTTP/2.0 stream requests which the server has received but not started or finishing responding to. Active HTTP/2.0 streams are streams for which the server is in the process of and has not finished sending header and data bytes. A server can have pending requests and no active HTTP/2.0 streams if it has pending requests to which it did not start responding.

\subsubsection{Pipelining Segment} a set of consecutive bytes sent by the server on a particular TCP stream containing header and data bytes for one or more HTTP/2.0 streams and that a) begins when a server with no prior pending requests starts responding to a new request and b) ends when the server sends the last bytes of all active HTTP/2.0 streams and has no pending requests. 

\subsubsection{Multiplexing Segment} a set of data bytes of a Pipelining Segment that are sent when the server has two or more active HTTP/2.0 streams. Multiplexing segments start and finish when the number of active HTTP/2.0 streams changes, i.e. after sending the last bytes of an active HTTP/2.0 stream or upon sending the first bytes of an active HTTP/2.0 stream. The number of objects in a multiplexing segment is defined as the number of active HTTP/2.0 streams in that segment. 

\subsubsection{Multiplexing Record} a TLS record that contains two or more HTTP/2.0 frame segments from different HTTP/2.0 streams. This occurs typically with small, HTTP/2.0 signaling and header frames.

\subsection{NPO Network Privacy Observatory Indicators}

\subsubsection{HTTP/2.0 / Encrypted HTTP} Proportion of the sum of TLS record content lengths for TLS records carrying HTTP/2.0 frames to the sum of all TLS record content lengths. 

\subsubsection{Pipelining / HTTP/2.0} Proportion of the number of bytes in pipelining segments to the sum of TLS record content lengths for TLS records carrying HTTP/2.0 frames.

\subsubsection{Multiplexing / Pipelining} Proportion of the number of bytes in multiplexing segments to the number of bytes pipelining segments.

\section{Observing Pipelining and Multiplexing Daily}
\label{sec:observing}
\begin{figure}[h!]
\begin{center}
\includegraphics[width=0.5 \textwidth]{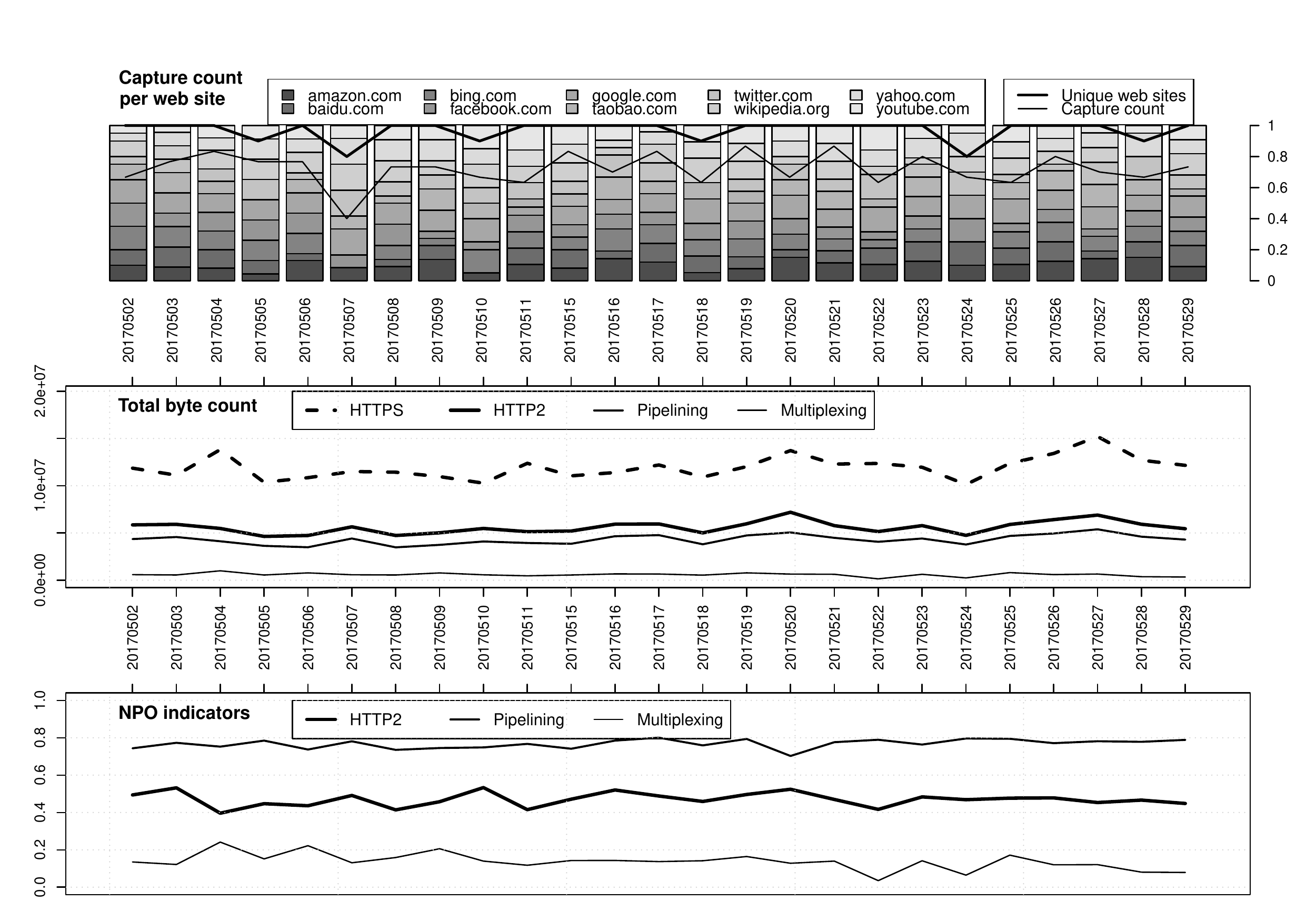} 
\end{center}
\caption{Evolution of NPO Indicators in May 2017. Top: number of unique web sites captured in the day in proportion to the 10 target web sites (thick line), number captures per day in proportion to 3x10 attempted captures (thin line), proportion of captures per web site (boxes). Center: sum for all web sites of per web site average byte count. 
Bottom: daily NPO indicators. In some days of the month we were not able to successfully capture any data.}
\label{npoindicatorstimeline}
\end{figure}

We have been observing multiplexing and pipelining daily on a small set of web site pages. The evolution of the NPO indicators for the set of web page sites is shown in Figure \ref{npoindicatorstimeline}. We can see that the NPO indicators are relatively stable and do not show a tendency to increase or decrease during the month, with HTTP/2.0 at one half, Pipelining at $4/5^{ths}$, and Multiplexing at $1/5^{th}$. Small variation may be caused by changes in actual number of captures performed per web site, deviating from target of 3 daily captures per each of the 10 web sites. The May 7 capture in particular has a significant drop in unique web sites (8/10) and captures (12/30), yet not visibly impacting the NPO indicators more than in other days. 

\begin{figure}[h!]
\begin{center}
\includegraphics[width=0.5 \textwidth]{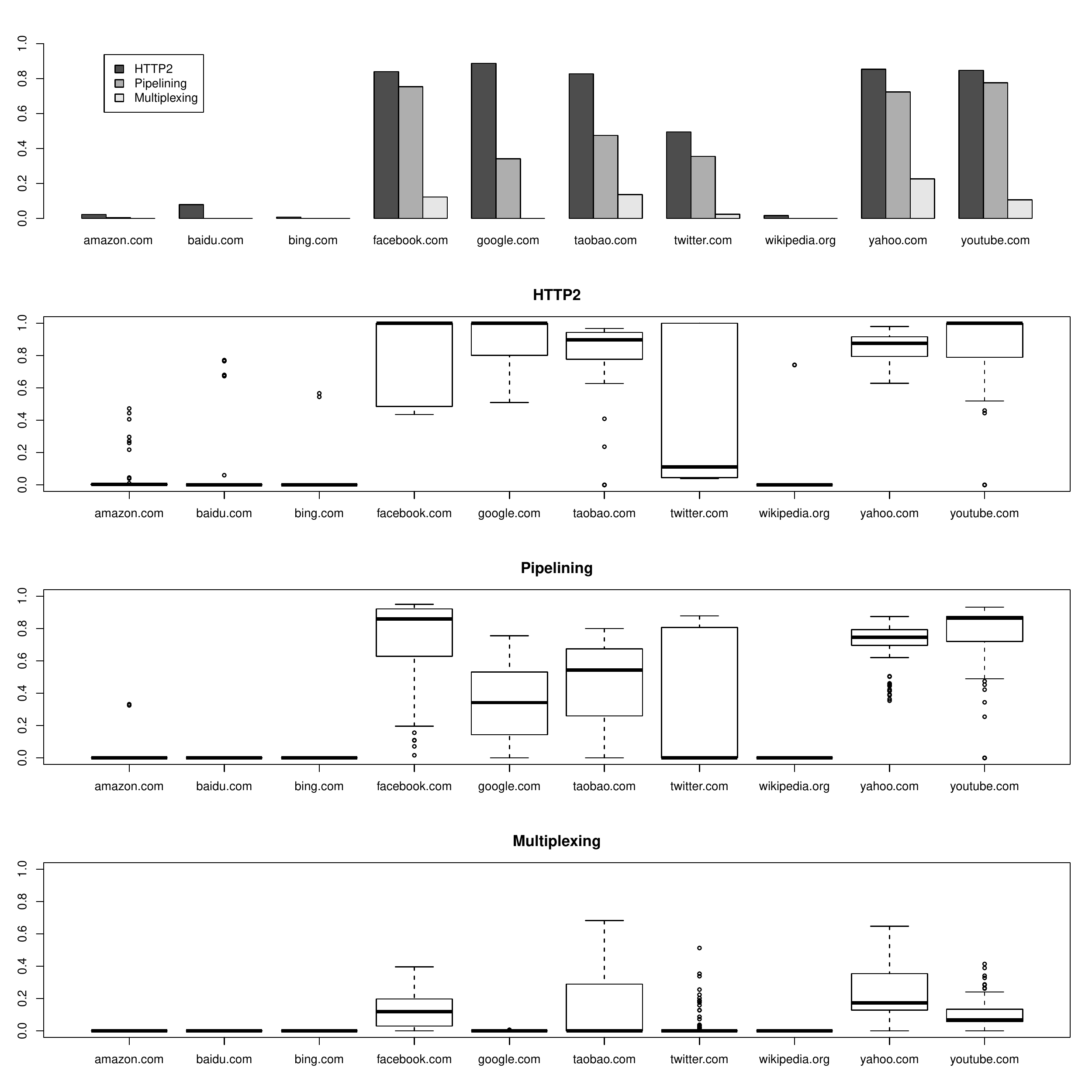} 
\end{center}
\caption{NPO Indicators per web site: average (top) and boxplot diagrams (bottom three).}
\label{npoindicatorsperwebsite}
\end{figure}

These results and in particular the pipelining results seem promising. However, when we split the data by website, we see that the results are not equally distributed per web site. The average NPO indicators for each web site and their boxplots are shown in Figure \ref{npoindicatorsperwebsite}. We can observe that 4 web sites have small values for the three indicators, meaning they are not using HTTP/2.0 greatly; and in the cases where they are, estimating web object size is not harder than in HTTP1/1. One web site does not use HTTP/2.0 in full, its HTTP/2.0 boxplot showing a close to zero median and almost 0-to-1 interquartile range. The remaining 5 web sites more fully utilize HTTP/2.0, with smaller interquartile ranges and larger than 80\% median. Looking at the Pipelining indicator for these 5 web sites, we see that only three have larger than 75\% median and that for the other two pipelining is only at around half. When we look at the Multiplexing indicator the landscape is dire: no web site has median larger than 20\%.

\begin{figure}[h!]
\begin{center}
\includegraphics[width=0.5 \textwidth]{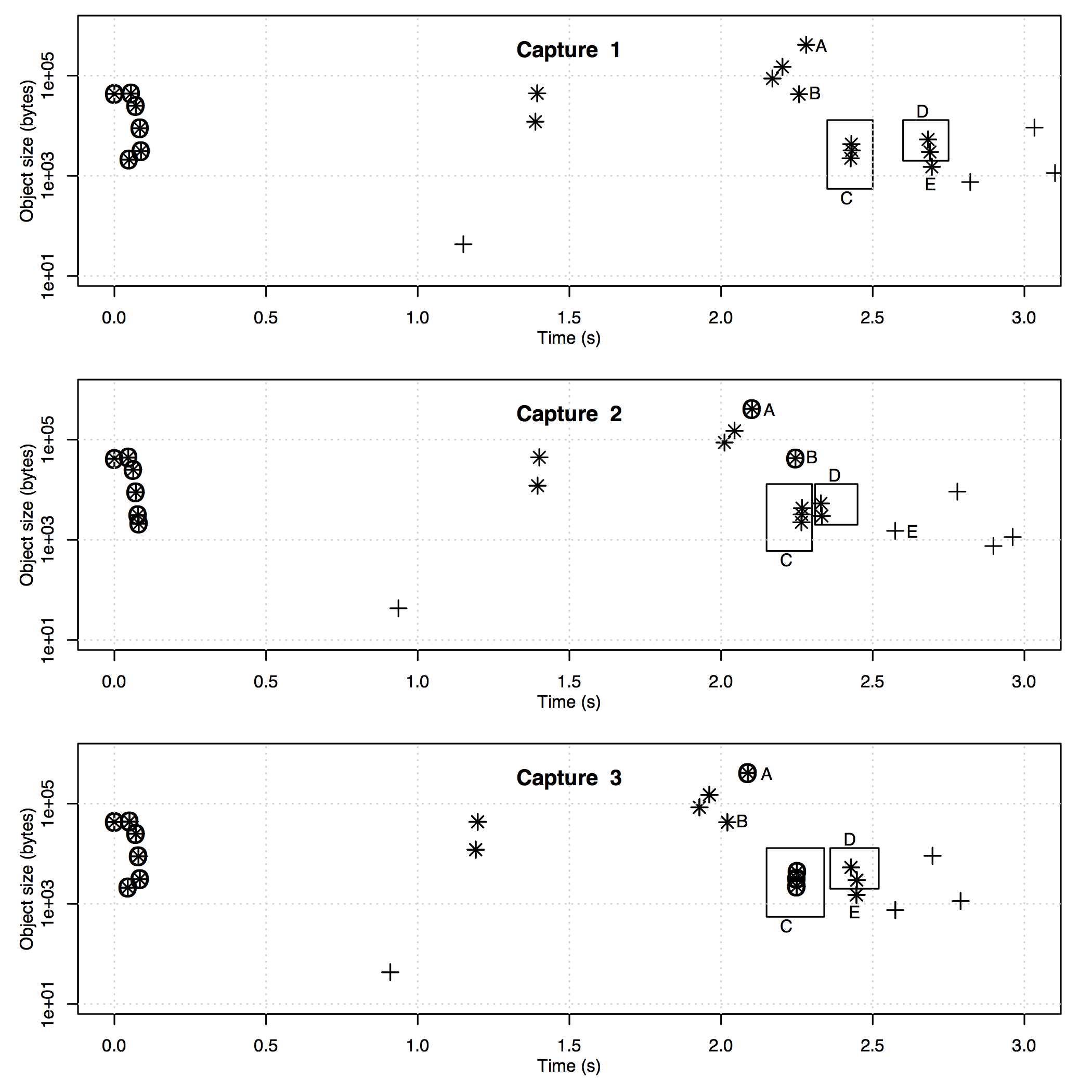} 
\end{center}
\caption{Web object size, time, pipelining, and multiplexing for three TCP streams of three different captures of the same web site
and that appear to carry the same web objects. Web objects are marked with asterisk if they are pipelined, asterisk and circle if they are multiplexed, and a plus sign if they're not pipelined or multiplexed. Web objects A, B, and E as well as groups C and D are marked out in the three captures.}
\label{example-tcp}
\end{figure}

To illustrate the variable nature of pipelining and multiplexing, in Figure \ref{example-tcp} we show three diagrams of web objects, their sizes, and the relative time at which they are sent. Each diagram represents the web objects in a TCP stream that we found in each of three captures of the same web site. The first observation is that although the sizes and timings of the web objects are not exactly the same, they are very similar and this part of the web site is likely to be relatively static. The second observation is that although most objects are either pipelined, multiplexed, or not multiplexed or pipelinined in all the three captures, some are not. If we intuitively map web objects between the TCP streams of the three captures and highlight objects A, B, and E, and groups C and D as shown in Figure \ref{example-tcp}, this is more evident. Object E is pipelined in captures 1 and 3 and not pipelined in capture 2, perhaps given the proximity in time to the objects in group D in captures 1 and 3. Arguments for why some objects are pipelined in some captures and multiplexed in others end there. Object A is multiplexed in capture 3 but not capture 1, when neighboring objects (B, C, and the two pipelined objects before A) do not change much between these captures. In capture 3 objects A and C seem to be multiplexed together, while in capture 2 this seems to happen with A and B.

\section{Experiment Methodology}
\label{sec:methodology}

The basis for the remaining of this work is a set of HTTPS requests issued on January 2017 to the first 10k of Alexa's top 1M web site pages \cite{durumeric_analysis_2013}. We prepend the "https://" prefix to each of the 10k Alexa's web site page names and load this URL in the browser. For each web site page we collect a tcpdump capture for all incoming and outgoing traffic through the client's Ethernet interface.   We also save a file with the pre-master secrets from the browser on each capture, which we then use to decrypt TLS records in the tcpdump pcap capture files. This is critical for the validation of our work as we can know exactly which part of the HTTP/2.0 streams each TLS record carries. Because of the complexity of web site pages, the HTTPS request for each of the 10k web sites will likely cause the browser to issue other HTTPS and non-HTTPS requests to the same and other server IP addresses. We use the pre-master secrets to parse the pcap files and populate a database with TLS records, HTTP/2.0 frames, and HTTP/2.0 streams. With this information we are able to characterize the extent of HTTP/2.0 pipelining and multiplexing and assess the success of an attack.

We use tcpdump 4.7.4 to capture pcap traffic files, tshark 2.2.5 to parse pcap files, and Chromium 57.0 on Ubuntu 16.04 desktop Linux with Selenium python webdriver 3.3.3 to open web pages.

\section{Web Traffic Characterization}
\label{sec:traffic-char}

Approximately 30\% of traffic bytes from our data set are not TLS encrypted and can be analyzed directly. 40\% of the traffic bytes are from prior versions of HTTP/2.0, which do not fully support pipelining and multiplexing and that mostly follow a simple sequential request-reply protocol that is straightforward to attack. Only 30\% of the traffic bytes are used for HTTP/2.0 and thus potentially offer pipelining and multiplexing protection to response size attacks. 

\subsection{HTTP/2.0 Server IP Addresses, TCP Streams, Web Objects}

We counted 75k TCP streams that send HTTP/2.0 frames for 145k HTTP/2.0 web objects from 2.8k distinct server IP addresses. The distributions of the number of TCP streams per distinct server IP address for the whole experiment is heavy-tailed. 50\% of the distinct server IP addresses have only one or two TCP streams each, whereas the top 1\% of server IP addresses have more than 42\% of the 75k TCP streams. The largest number of TCP streams per server IP address is 2334. This may relate to the popularity of the content available in some servers throughout the set of 10k web sites for which we captured traffic. 

We also found that one third of the TCP streams do not carry any data or header frames but only HTTP/2.0 signalling frames such as SETTINGS or WINDOW\_UPDATE. Only 49k TCP streams from 2.5k distinct server IP addresses actually carry web objects. The largest number of web objects downloaded from a single TCP connection is 541, the top 1\% of TCP streams have 19\% of the web objects, and more than 60\% of the TCP streams have only one web object, totaling 18\% of the web objects. This means that 18\% of the web objects carried in the 30\% of traffic bytes that fully support pipelining and multiplexing are not multiplexed or pipelined - as they do not share a TCP stream with other web objects.

\subsection{Web Object Size and HTTP/2.0 Frame Size}

\begin{figure}[h!]
\begin{center}
\includegraphics[width=0.3 \textwidth]{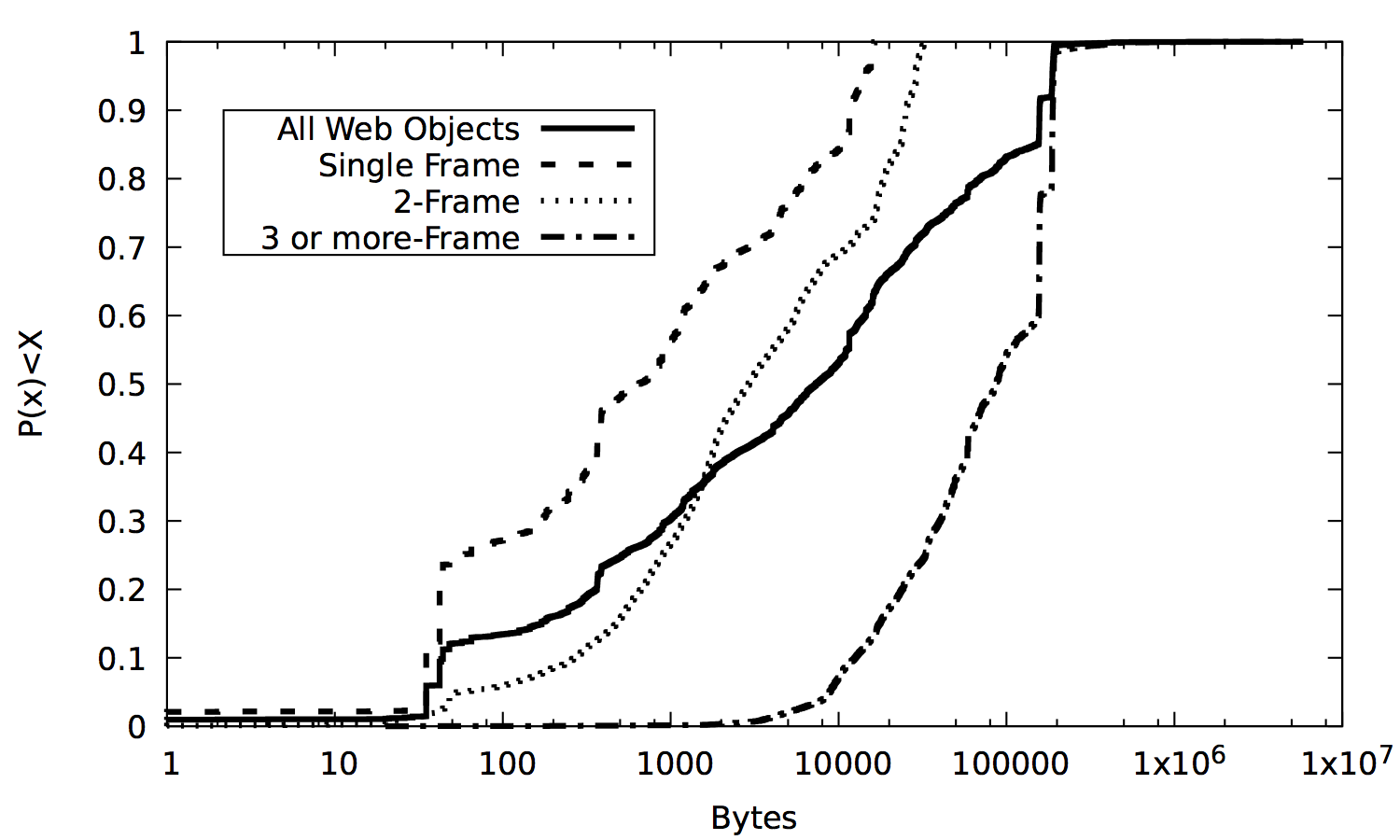} 
\end{center}
\caption{Distribution of the sizes of web objects.}
\label{webobjsizeframescdf}
\end{figure}

Figure \ref{webobjsizeframescdf} shows the distribution of web object sizes and how this changes with the number of frames per object. 45\% of the web objects are delivered in a single HTTP/2.0 frame, 18\% in two HTTP/2.0 frames, and the remaining 37\% in three or more HTTP/2.0 frames. As expected, the single-frame web object sizes tend to be smaller than two-frame and three or more-frame web objects. However we notice that 1) half of the single frame web objects are larger than 1k bytes, some reaching 10k  bytes and 2) more than 20\% web objects are sent in two frames even if their size is relatively small, below 1k bytes. 

\begin{figure}[h!]
\begin{center}
\includegraphics[width=0.3 \textwidth]{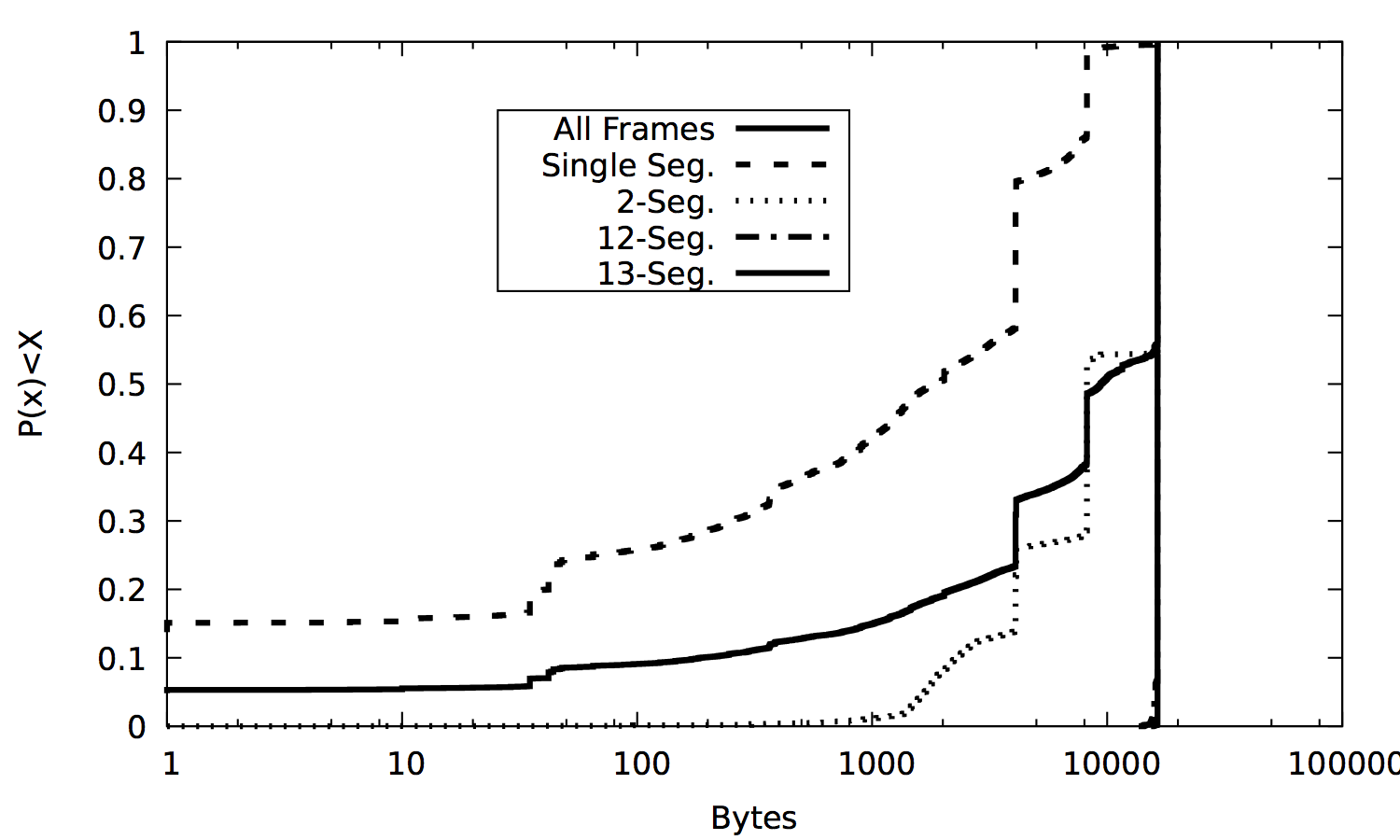} 
\end{center}
\caption{Distribution of the sizes of HTTP/2.0 response data  frames.}
\label{framesizecdf}
\end{figure}

Figure \ref{framesizecdf} shows the distribution of HTTP/2.0 frame sizes and how this changes with the number of segments per frame. 35\% of the frames are delivered in a single HTTP/2.0 frame segment, 15\% in two HTTP/2.0 frame segments, 22\% in 12 frame segments, and 14\% in 13 HTTP/2.0 frame segments. The largest observed frame size is 16384 bytes, which according to the standard corresponds to the smallest possible maximum frame size setting for HTTP/2.0~\cite{belshe2015hypertext}.

\subsection{TLS records and HTTP/2.0 Frame Segments}

\begin{figure}[h!]
\begin{center}
\includegraphics[width=0.3 \textwidth]{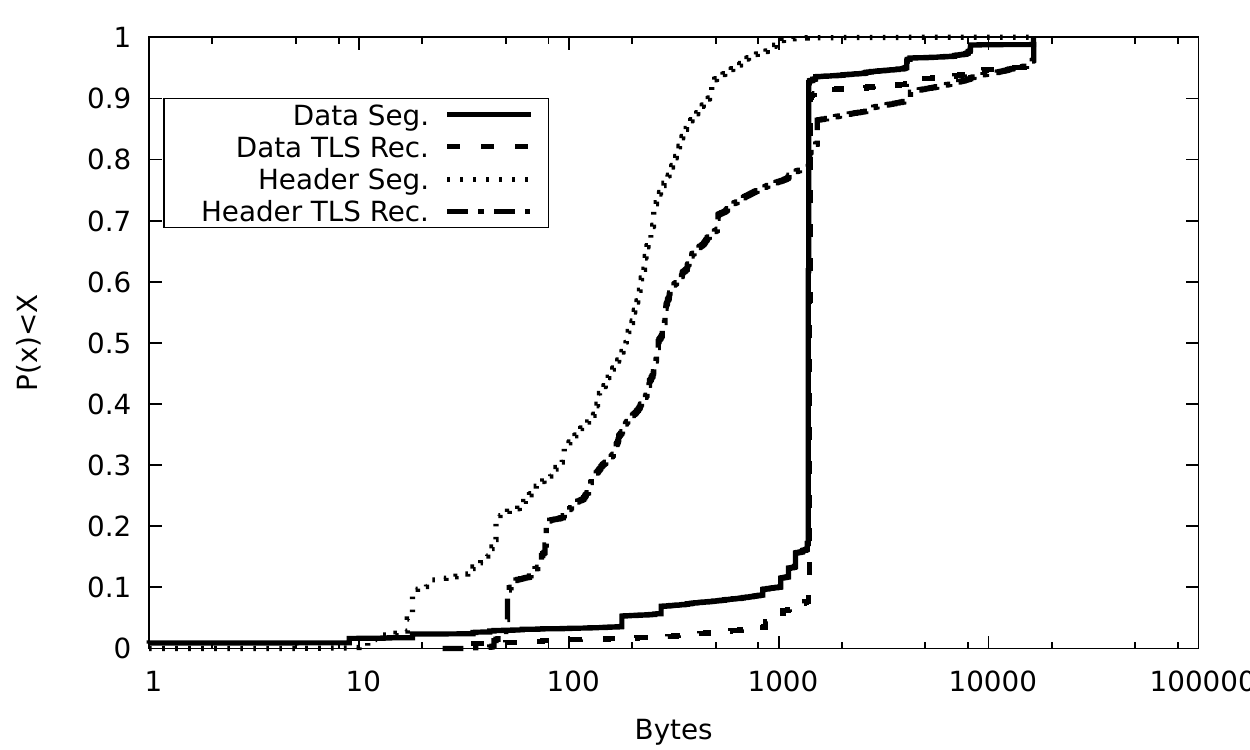} 
\end{center}
\caption{Distribution of the sizes of data and header server-to-client HTTP/2.0 frame segments and TLS application records.}
\label{framesegtlsrecsizecdf}
\end{figure}

We collected over 8.6 million TLS records, from which 4.1 million are HTTP/2.0 application records and 3.7 million HTTP/2.0 application records sent by the server. We observe that most TLS records are either used to send data frames or non-data frames. In fact, only 1\% of server-to-client TLS records have both data and non-data frame segments. 3.29 million TLS records are used to send HTTP/2.0 data frames and 145k TLS records to send header HTTP/2.0 frames.  Figure \ref{framesegtlsrecsizecdf} shows the distribution of TLS record and HTTP frame segment sizes, for either data or header frames. Most data HTTP/2.0 frame segments are 1381 and 1389 bytes long, corresponding to TLS record sizes of 1405 bytes and 1413 bytes, respectively. This could be related to cross-layer interaction between IP MTU and TLS record size.

From the TLS application records that encapsulate server-to-client HTTP/2.0 data frames, 94\% encapsulate one HTTP/2.0 data frame segment. The difference between TLS record size and the size of the HTTP/2.0 frame segment it carries is 24 bytes for an overwhelming 96\% of data frame records, mostly due to the (TLS record, frame segment) size pairs (1381, 1405) and (1389, 1413) that take up 86\% of the single frame TLS records. There is also a 24 byte difference for 74\% of the other single frame TLS records. More than 99\% of the TLS records that encapsulate HTTP/2.0 header frames are single frame segment records and 97\% out of those have a 33 byte difference between TLS record size and HTTP frame segment size.

\section{Extent of Pipelining and Multiplexing}
\label{sec-extent}

\begin{figure}[h!]
\begin{center}
\includegraphics[width=0.3 \textwidth]{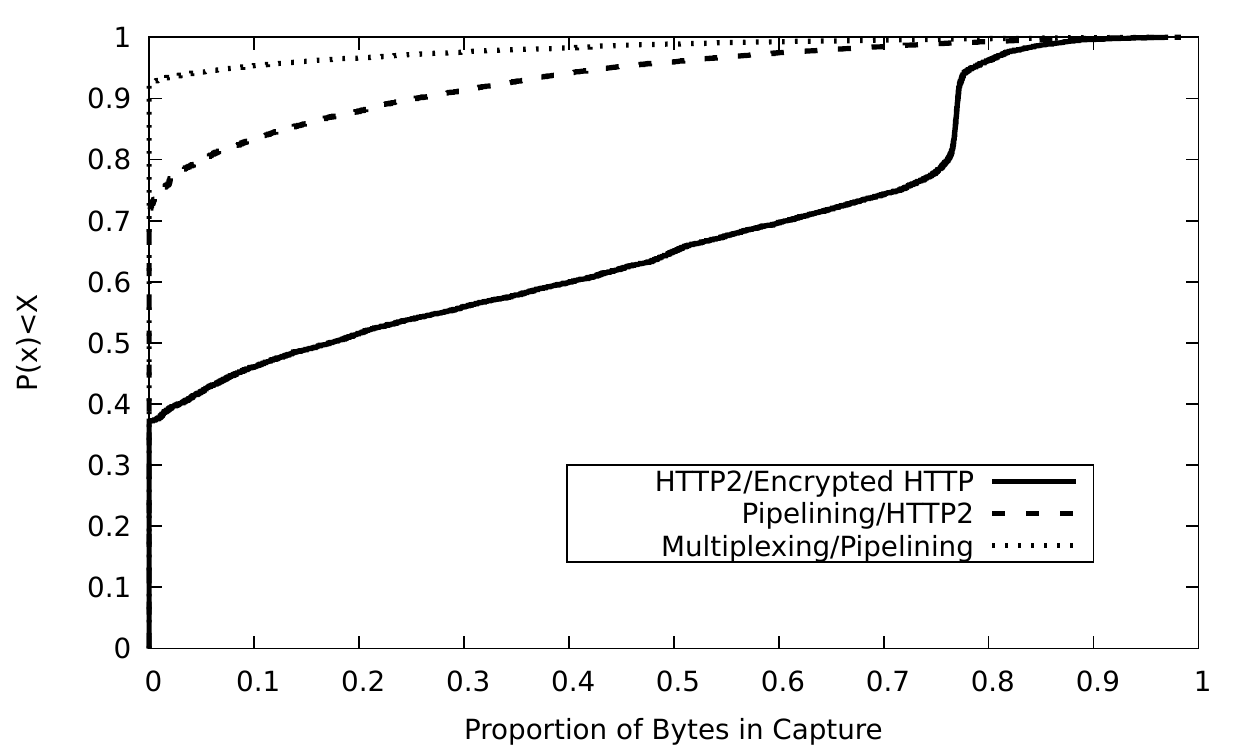} 
\end{center}
\caption{Proportion of bytes in each capture: 1) HTTP/2.0 bytes to total encrypted HTTP bytes, 2) Pipelining bytes to HTTP/2.0 bytes, and 3) Multiplexing bytes to Pipelining bytes.}
\label{bytecdf}
\end{figure}

Figure \ref{bytecdf} shows that the extent of HTTP/2.0, Pipelining, and Multiplexing usage is relatively small: 1) more than 35\% of web site captures have no HTTP/2.0 bytes, 2) more than 70\% have no pipelined bytes, and 3) more than 90\% have no multiplexed bytes. This means that for the more than 70\% of the captures (which do not have pipelined bytes) figuring out the size of all web objects is no more difficult than in encrypted HTTP/1.0. The extent of pipelining and multiplexing is foremost related to the adoption of HTTP/2.0. We expect the extent of pipelining and of multiplexing to increase as this adoption increases\footnote{\url{http://}isthewebhttp2yet.com}. Captures with strong ($>$ 75\%) pipelining or multiplexing effect are small in number: only 1\% of the captures for strong pipelining, 0.3\% for strong multiplexing, and 0.08\% for both strong pipelining and multiplexing. 

Overall, only 42.7k web objects were pipelined and 7.4k multiplexed, out of 145k. We found 110k pipelining segments in our data set, 92\% of which contain data for a single HTTP/2.0 stream, 4\% for two streams, and the remaining 4\% for 3 or more streams. The largest 60 pipelining segments have 20 or more HTTP/2.0 streams. We also found 9k multiplexing segments, half of which with two HTTP/2.0 streams. The largest 70 multiplexing segments have 20 or more HTTP/2.0 streams.

\subsection{Pipelining and the number of web objects per TCP stream}

Out of the 49k TCP streams that carry web objects, 29k carry only one web object. As pipelining requires at least two web objects, this immediately leaves out 29k web objects from pipelining. The question for these 29k web objects then becomes whether the web client is opening a new TCP connection to the same server IP address for each web object that it has to retrieve from that server or does it only request one web object from that server throughout the capture of the web site traffic. It turns out that 16.7k web objects could not have been pipelined as they are the only web object requested from their server IP address in the web site traffic capture. The remaining 12.3k could have been pipelined, depending on the specifics of how the web site redirects the web browser to subsequent HTTP requests and if the web client opened a single TCP connection for all objects in that server IP address. Additionally, 13.6k TCP streams carry two or more non-pipelined web objects, totaling 54.5k web objects that could also have been pipelined depending on the specifics of the web site.

\begin{figure}[h!]
\begin{center}
\includegraphics[width=0.3 \textwidth]{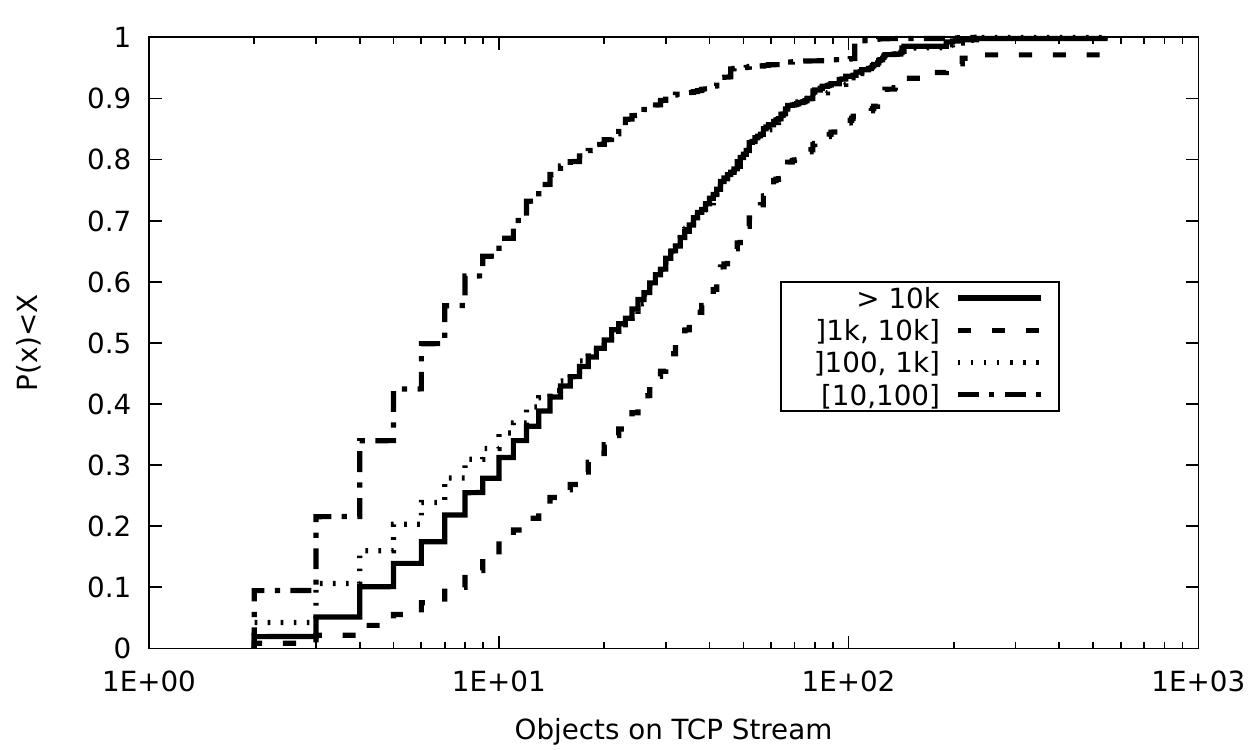} 
\end{center}
\caption{Distribution of the number of web objects in the same TCP stream of a web object of a given size.}
\label{perobjsize_countobjsintcpstream_pipe}
\end{figure}

We observe that the extent of pipelining is much stronger for web objects between 1k and 10k bytes than for other sizes. 48\% of 33k web objects with size in range ]1k, 10k] are pipelined compared to 26\% of 67k web objects larger than 10k, 25\% of 24k web objects with size in range ]100, 1k], and 18\% of 18k web objects with size in range ]10, 100]. Figure \ref{perobjsize_countobjsintcpstream_pipe} shows the number of web objects in the same TCP stream as a given object, grouped by that object's size. We can observe that in the ]1k, 10k] range, in which the extent of pipelining is stronger, the number of web objects per TCP stream is larger than for other ranges. This seems to support the following: the more web objects there are in a TCP stream, the stronger will the pipelining effect be. The precise timing of the object requests should of course also be of relevance.

\subsection{Multiplexing and the number of HTTP/2.0 frame segments per pipelined web object}

Out of 42.7k pipelined web objects, 14.3k have a single HTTP/2.0 frame segment. If by chance the single HTTP/2.0 frame segment objects are sent by the server in the middle of other, multi-HTTP/2.0 frame segment web objects, then they will be part of a multiplexing segment. This happens to only 1.7k of the single-HTTP/2.0 frame segment web objects.  

Out of the remaining 28.4k web objects with two or more HTTP/2.0 segments, 22.7k objects are not multiplexed. This is extremely interesting as there does not appear to be any obvious reason why pipelined web objects with more than one HTTP/2.0 frame segment should not be multiplexed. Having received requests for two or more web objects in the pipelining segment and with more than two HTTP/2.0 frame segments to send, the server should be able to round robin the frame segments and multiplex the web objects. In a majority of cases (22.7k out of 28.4k) it does not. This contributes to the small extent of multiplexing.

The extent of multiplexing for smaller web objects is much smaller than for larger sizes. From 6\% for web objects larger than 10k and 8\% for web objects with size in range ]1k, 10k], multiplexing drops to 2\% for web objects with size in range ]100, 1k] and to 0.3\% for range ]10,100]. This could be explained by the number of single HTTP/2.0 frame segment web objects in web object size ranges: 95\% of the web objects in size range ]10, 100] have a single HTTP/2.0 frame segment, 80\% in range ]100, 1k], 40\% in range ]1k, 10k], and only 4\% for web objects larger than 10k bytes. Smaller web objects seem to be multiplexed to smaller extent than others because they mostly have single HTTP/2.0 frame segment.

\section{Attack Assumptions and Worst Case Results}
\label{sec-attack-assumptions}

Worst case attack results can be characterized by defining the boundary for which the attacker cannot do worse in underestimating or overestimating web object size, according to some attack assumptions.  We define a set of attack assumptions and provide results for worst case attacks under each assumption. Our rationale is that 1) best case attacks would always be able to predict web object sizes accurately so there is no point in defining best case boundaries here and 2) if the resulting worst case boundary turns out to be good enough for the attacker, then the only thing that the attacker needs is to meet the assumption in order to be successful. We start by showing that even without pipelining estimating the size of web objects is not free of error and establish a no-pipelining baseline to better understand the impact of pipelining and multiplexing.

\subsection{Estimating Web Object Size Without Pipelining}
\label{sec:error}

For the 102k web objects that are not pipelined it is straightforward to estimate their size. For non-pipelined web objects, the client issues the request for the object and has to wait for the server to send the web object before it sends a new request. Thus we simply need to sum the sizes of server-to-client TLS records between client-to-server TLS records. However, this sum will almost never be exactly equal to the size of the web object: TLS records carrying the web object have encryption overhead and may include HTTP/2.0 signaling. 

\begin{figure}[h!]
\begin{center}
\includegraphics[width=0.3 \textwidth]{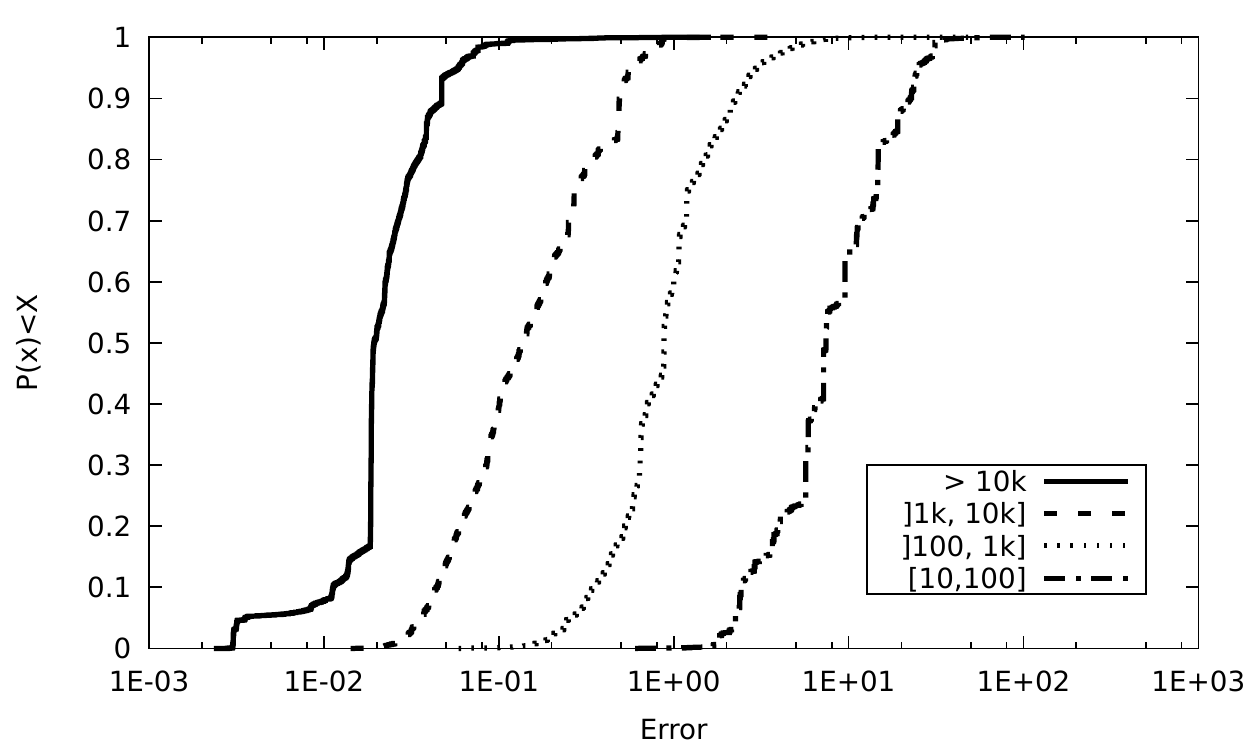} 
\end{center}
\caption{Cumulative distributions of non-pipelined web object estimate error for different ranges of web object sizes (in bytes)}
\label{estimate-nopipe}
\end{figure}

We take object size $s_{act}$ and estimated size $s_{est}$ as the sum of all TLS records that contain frame segments of that object and compute error $e = (s_{est}-s_{act}) / (s_{act})$. Figure \ref{estimate-nopipe} shows the distribution of error $e$ for different ranges of web object size $s_{act}$. Larger web objects have smaller relative error and very small (< 100 bytes) web objects have extremely large error. Objects smaller than 10 bytes have all error $e$ larger than 10.

\subsection{Attack Assumptions}

\subsubsection*{Assumption 1} The positions of the beginning and end bytes of all pipelining segments are known, as well as the number of web objects in each pipelining segment.

\subsubsection*{Assumption 2} The positions of the beginning and end bytes of all multiplexing segments are known, as well as the number of web objects in each multiplexing segment.  

\subsubsection*{Assumption 3} The TLS records that carry each HTTP/2.0 stream are known.

Assumption 3 is the strongest assumption and covers pipelining and multiplexing segments. Assumption 2 covers pipelining segments only. Assumption 1 is the weakest assumption, only covering client-server requests and not pipelining or multiplexing segments. Multiplexing records are not covered by any of the assumptions and  we believe this would require breaking the encoding on the TLS records. 

Our three assumptions yield high and low value estimates. The worst case estimates $s_{est}$ are computed using the high or low value estimate that yields the largest error $e$. In each of the assumptions and in the case where two or more web objects are in the same segment (or record in the case of assumption 3), we additionally assume that all web objects in the segment/record are extremely small except for one, which is dominant and has approximately the size of the pipelining segment. Since the exact dominant web object is not assumed to be known, the high value estimate for all web objects in the segment/record is the size of the segment/record and the low value estimate is zero. In the case of single web object segment/record, both the high and low value estimates are equal to the size of the segment/record.

For assumptions 1 and 3 we additionally require that the request header size is known so that it can yield a value for $s_{est}$ that considers only the data part of the server response. 

\subsection{Worst Case Results}

\begin{figure}[h!]
\begin{center}
\includegraphics[width=0.3 \textwidth]{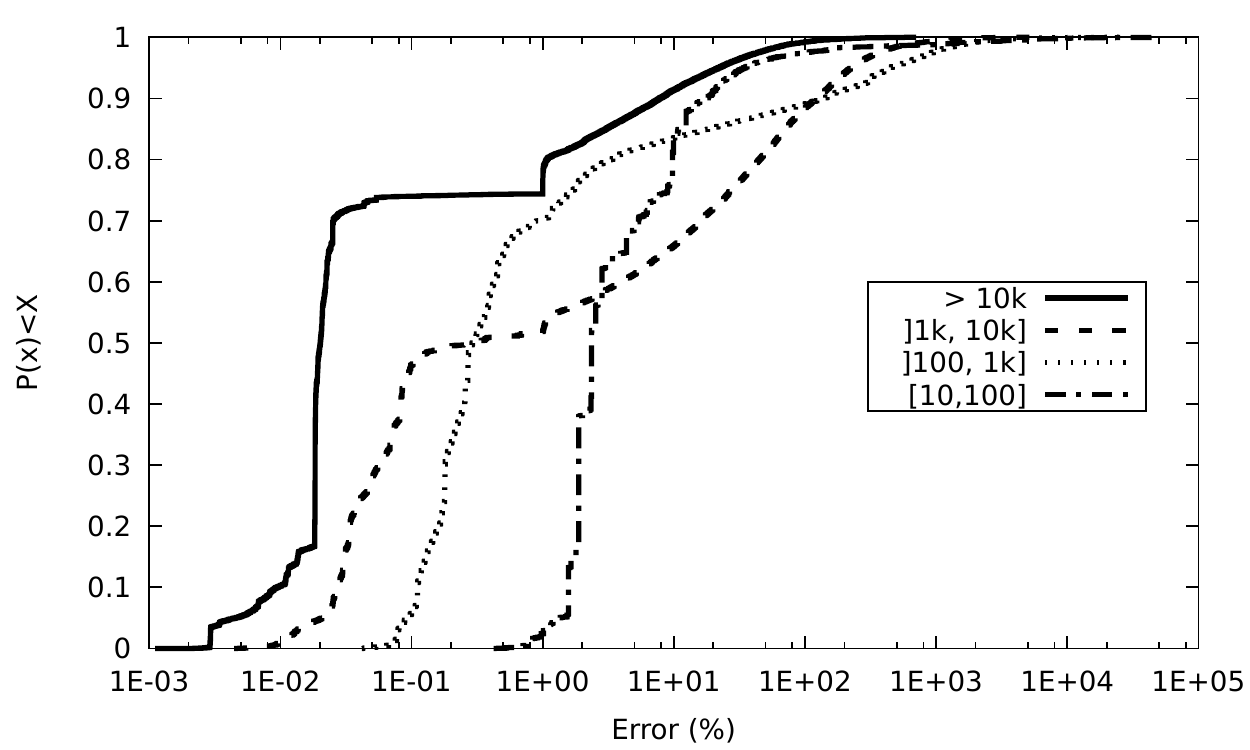} 
\includegraphics[width=0.3 \textwidth]{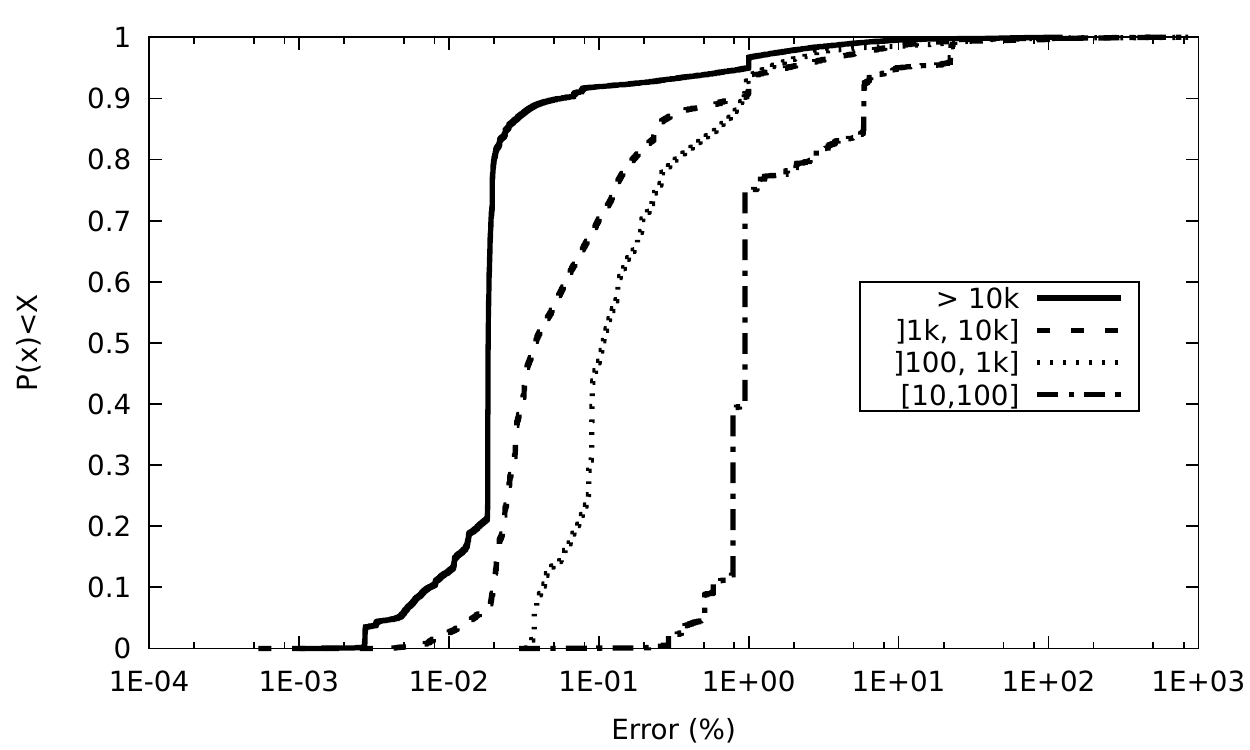} 
\includegraphics[width=0.3 \textwidth]{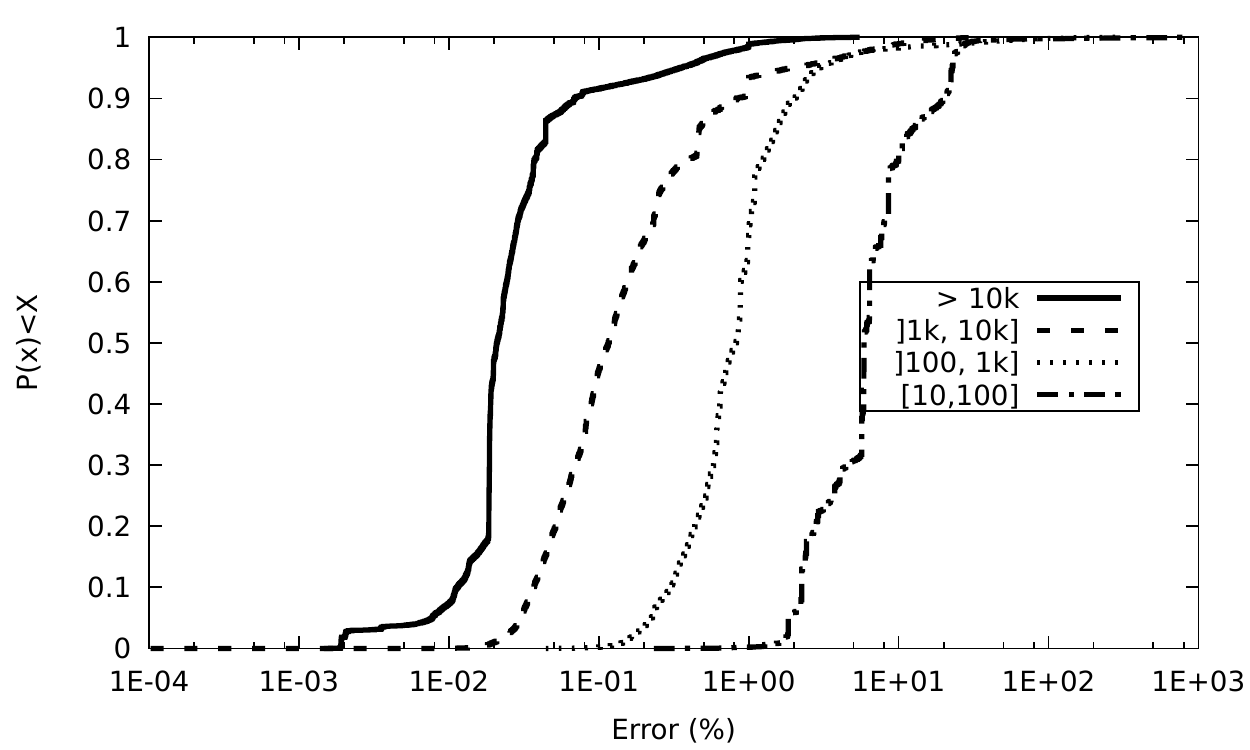} 
\end{center}
\caption{Error of worst case assumptions, broken down by assumption: A1, A2, A3 from top to bottom.}
\label{estimate-a-star}
\end{figure}

\begin{figure}[h!]
\begin{center}
\includegraphics[width=0.3 \textwidth]{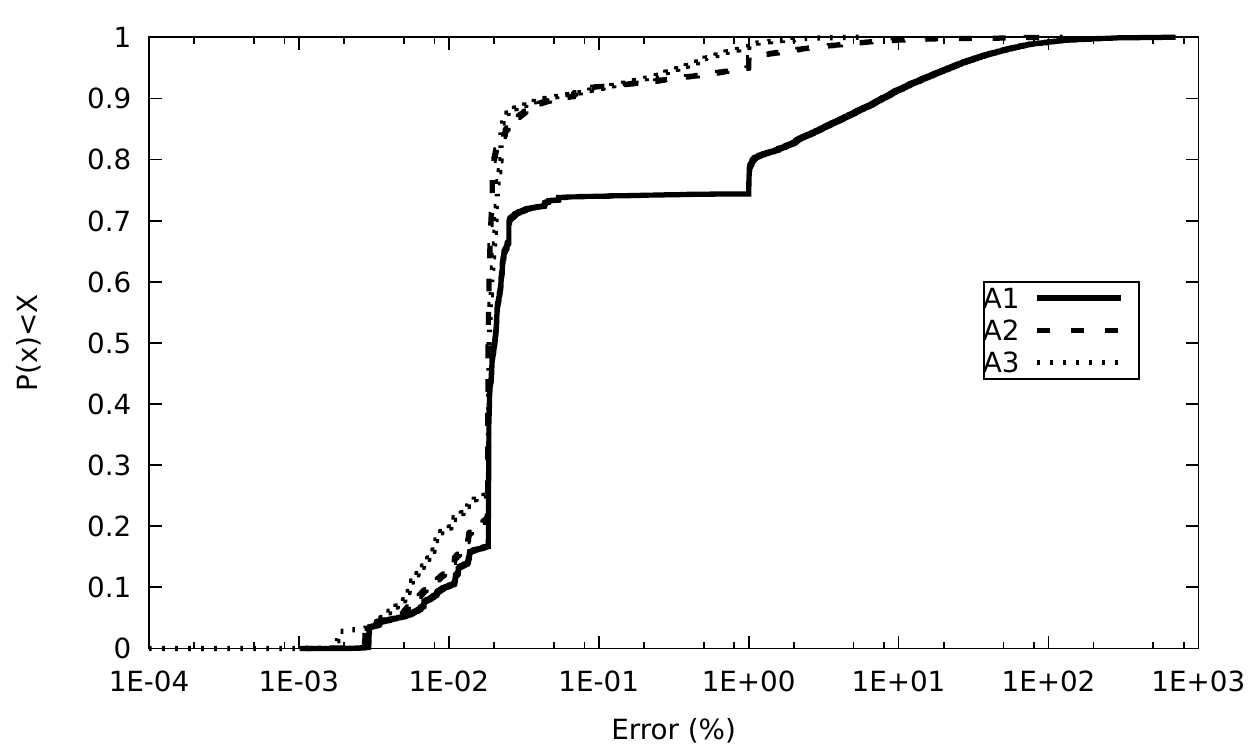} 
\includegraphics[width=0.3 \textwidth]{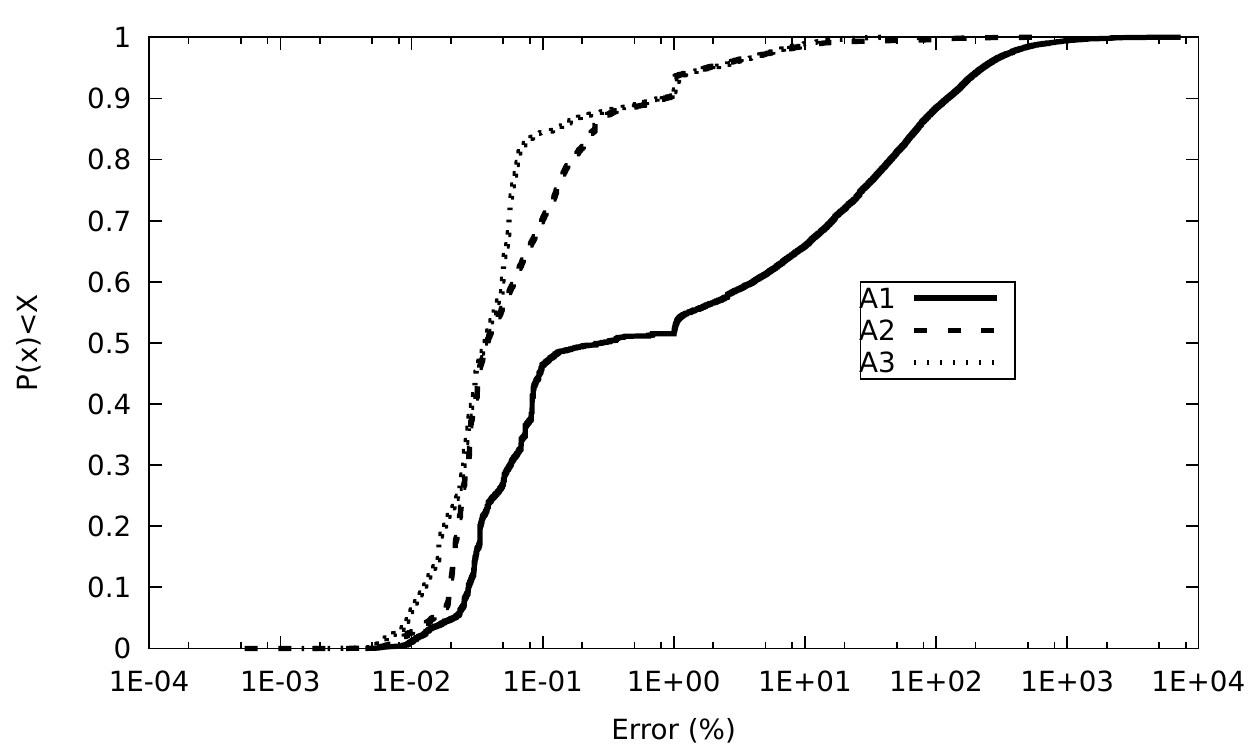} 
\includegraphics[width=0.3 \textwidth]{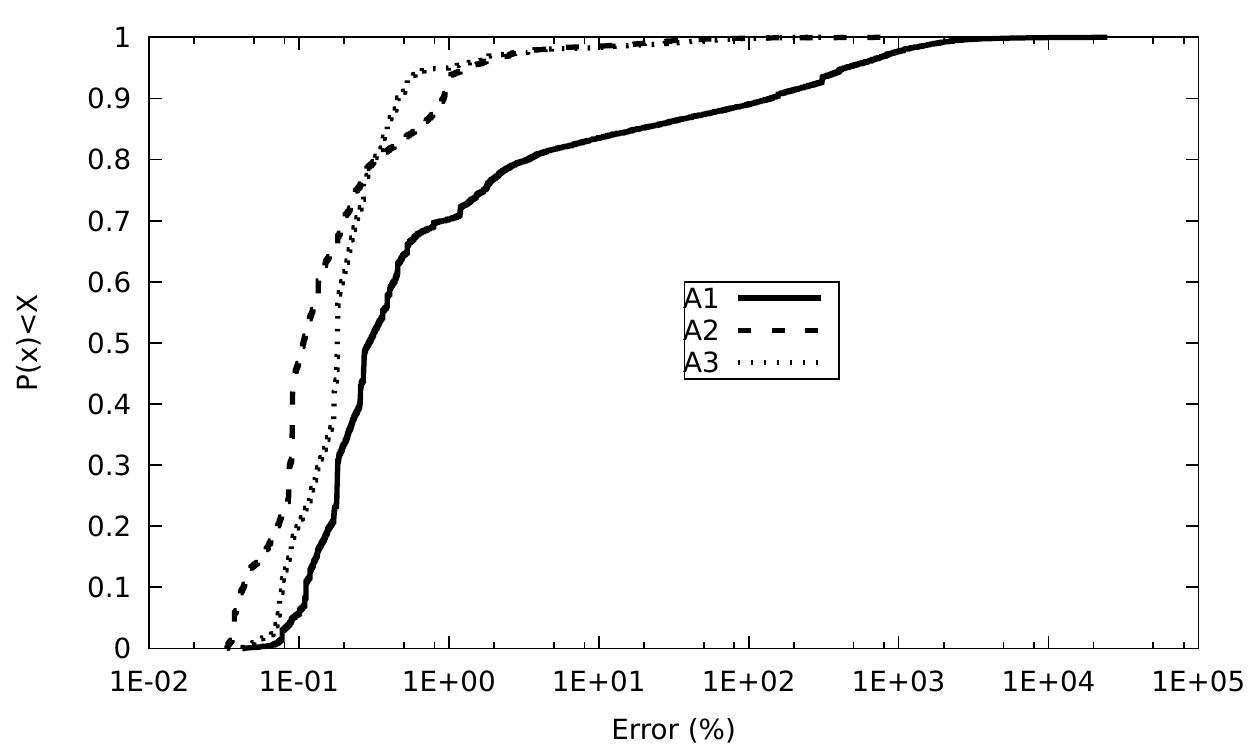} 
\includegraphics[width=0.3 \textwidth]{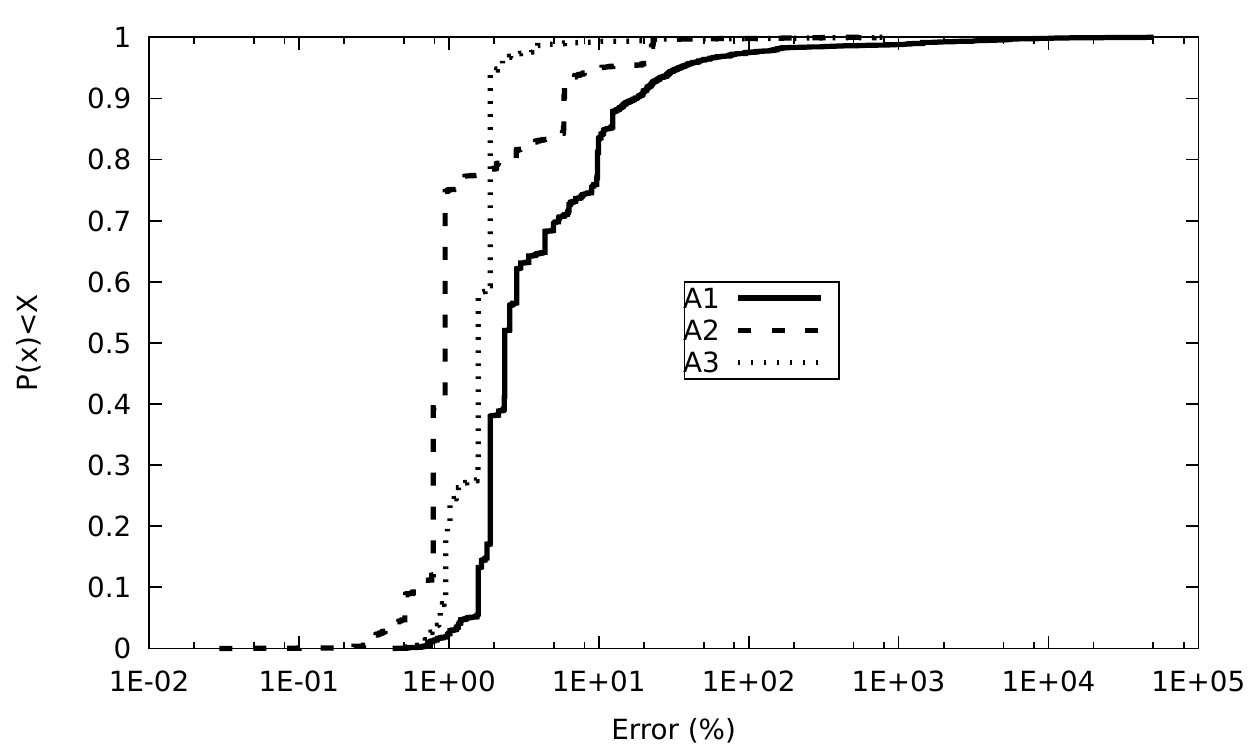} 
\end{center}
\caption{Error of worst case assumptions, broken down by web object size: > 10k , ]1k,10k], ]100,1k], ]10,100] from top to bottom.}
\label{estimate-a123-star}
\end{figure}

We provide results per assumption broken down by web object size in Figure \ref{estimate-a-star} and per web object size broken down by assumption in Figure \ref{estimate-a123-star}.

We can see from Figure \ref{estimate-a-star} that in general and for the different assumptions, larger web objects have smaller relative error. The impact of the mechanisms that can affect estimated web object size (like pipelining and multiplexing segments, multiplexing TLS records, and TLS encoding overhead) seems not to grow as much as the web object size. Figure \ref{estimate-nopipe} shows the error for non-pipelined web objects and also supports this intuition. One exception is the case of web objects between 1k and 10k under assumption A1. The worse error distribution in this size range is likely due to the proportion of pipelined web objects, which in this range is almost twice as much than in other ranges.

Figure \ref{estimate-a123-star} shows that the error is relatively small under weaker assumption A1 and for larger web objects. Under assumption A1, 70\% of web objects larger than 10k have an error smaller than 2\%, while 50\% of web objects between 1k and 10k have an error smaller than 10\%. Figure \ref{estimate-a123-star} also shows that A2 and A3 curves are very close to each other. An attacker would thus not be required to meet stronger A3 assumption but only A2 to have worst case results similar to A3. Overall we observe a more extensive use of pipelining and multiplexing that can bring the error curves closer to the bottom right part of the CDFs, especially for 1k and larger web objects. 

\subsection{Results for Pipelining and Multiplexing Objects Only}

\begin{figure}[h!]
\begin{center}
\includegraphics[width=0.3 \textwidth]{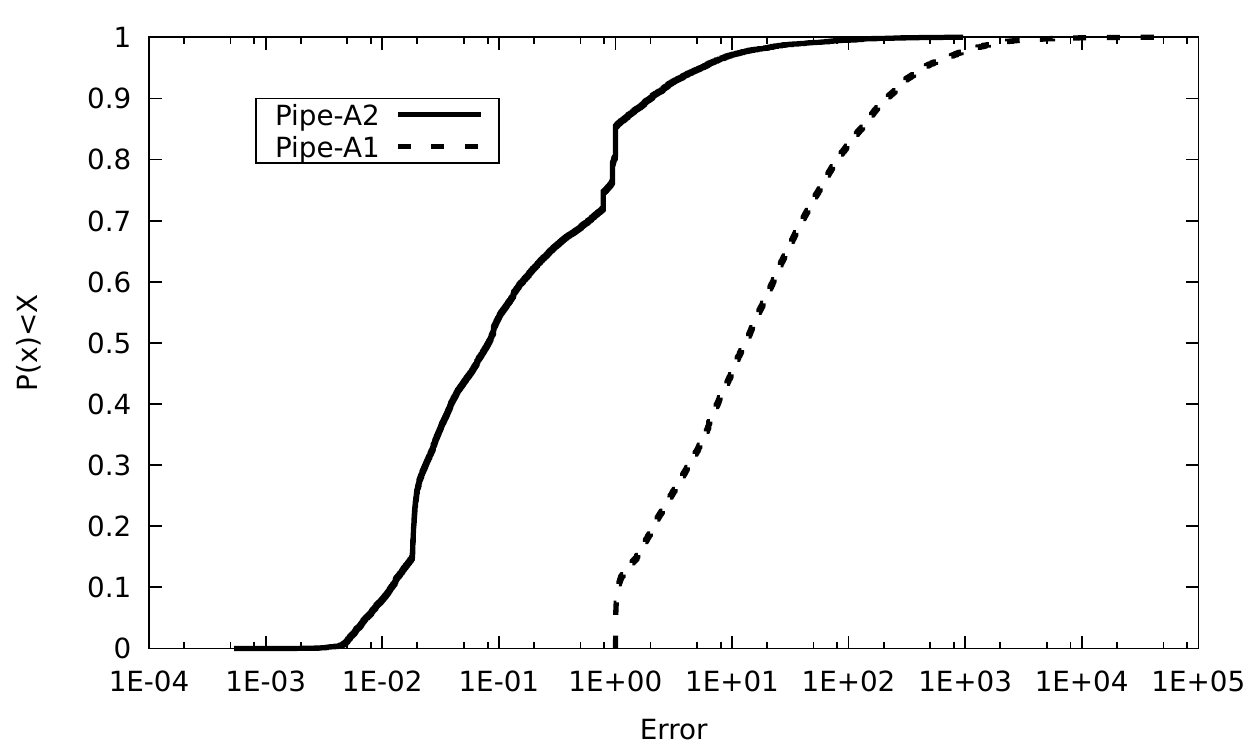} 
\includegraphics[width=0.3 \textwidth]{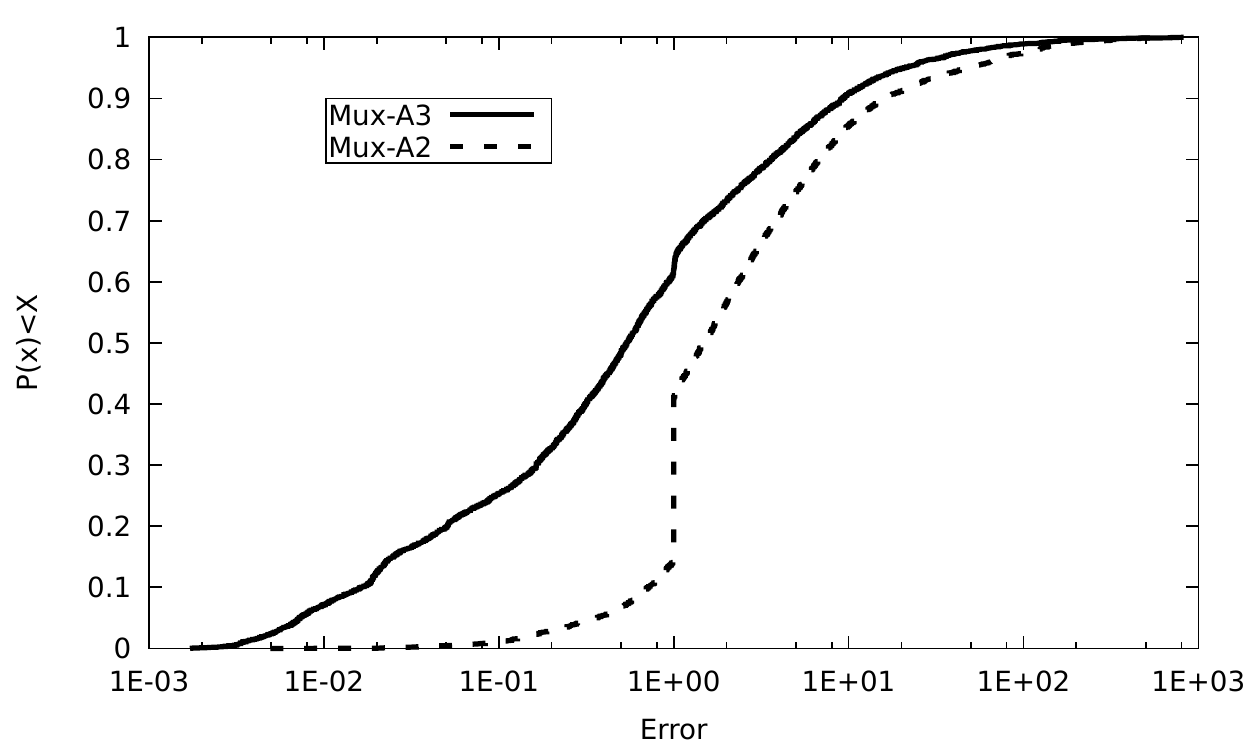} 
\end{center}
\caption{Comparison of estimation error $e$ for pipelined (top) and multiplexed (bottom) web objects. For pipelined web objects we compare assumption A2 which covers pipelining with weaker assumption A1 which only covers client-server requests. For multiplexed web objects we compare assumption A3 which covers multiplexing with weaker assumption A2 which only covers pipelining.}
\label{worstcase-pipe+mux}
\end{figure}

The previous section provides worst case results for non-pipelined, pipelined, and multiplexed objects together under the different attack assumptions. Here we focus on pipelined and multiplexed web objects only, in order to better gauge the effect of pipelining and multiplexing. Figure \ref{worstcase-pipe+mux} shows how strong the pipelining and multiplexing effects are by comparing worst case results for the weaker assumption that covers the effect (A2 for pipelining and A3 for multiplexing) with the strongest assumption that does not cover the effect (A1 for pipelining and A2 for multiplexing). 

\section{Example Attack}
\label{sec-example-attack}

In the previous section we defined attack assumptions and discussed worst case results without proposing a specific attack. In this section we provide an illustrative example of an attack that meets a weaker form of Assumption 2 in which the number of web objects in each segment is not known. This attack, however, is only valid for some server IP addresses.

\subsection{Description}

Our example attack has two parts. In the first part we address the feature in HTTP/1.1 and beyond that allows a TCP connection to be reused for multiple requests. We segment the TLS records of each TCP connection into sets of HTTP request-response sequences that are contained entirely in each segment. Our assumption in segmenting TCP connections is that the TLS records of ready-to-send HTTP responses are sent back-to-back. We look for gaps in the sequence of timestamps of these records and use these gaps to segment the TCP connections.  In HTTP/1.1 a gap will exist between the timestamp of the last TLS record of the current response and the timestampt of the first TLS record of the next response. This gap is at least one round trip time, as the client has to wait until the the current response is completely received before sending the next request. More generically in order to include pipelining and response multiplexing, we take a gap in the back-to-back TLS record sequences from the server as indication that all requests sent by the client and received by the server up to that point have been served. We declare a gap in the TCP connection when the difference between the timestamps of consecutive TLS records from the server is 1) larger than 0.5 seconds or 2) larger than 20 times the average back-to-back response gap from the server, normalized to a 1500 byte TLS record size. 

In the second part we analyze the sets of HTTP request-response of the segmented TCP connections to compute response sizes. A basic approach would not consider pipelining and multiplexing and simply take the sum of all TLS record sizes from the server on each segment of the TCP connection as an estimate of HTTP response size. The approach that considers pipelining to estimate HTTP response sizes relies on our intuition on the following three side-channel information that leak from TLS. 1) HTTP/2.0 signaling is visible through small-sized TLS records (less than 60 bytes) that indicate the beginning of an HTTPS connection or the beginning or end of an HTTP response. 2) Large HTTP responses generate back-to-back TLS records with relatively large size corresponding to network ($\sim$ 1.5 kB) or TLS ($\sim$ 16 kB) MTUs. 3) Request and response headers are sent in their own TLS records, typically yielding record sizes smaller than the MTU. We use this information to estimate the start and finish times for HTTP responses. With start and finish times it is possible to identify segments in the TLS record size sequences that contain entire responses -- some of which will be pipelined but not multiplexed while others will be entirely or partially multiplexed. We sum all TLS record sizes to get an estimate of non-multiplexed responses and ignore multiplexed responses.

In a previous experiment \cite{morla_initial_2016} we found  that HTTP/2.0 connections from the same type of server IP addresses have the a same initial sequence of server-to-client TLS record sizes. After  inspecting captures of one of these types of server IP address, we decided to use the following information for our attack:

\begin{itemize}
\item HTTP request and response headers are sent in TLS records with sizes ranging from 70 to 350 bytes.
\item 41 byte-length TLS records from the server indicate that a response has finished.
\end{itemize}

The output of the estimation process is a list of sizes and start and finish positions of estimated responses per TCP stream from each web server.

\subsection{Target Characterization}

Our attack is only suitable for some server IP addresses. We found 47.7k target TCP streams from 388 unique server IP addresses that match our attack target's initial sequence of TLS record sizes. From these only 33.9k TCP streams actually carry web objects - a total of 87.9k web objects, from which 10.8k are pipelined and 0.5k are multiplexed. Like in the case of the whole data set, attack target web objects larger than 10k bytes are the largest group with 53\% of the web objects; 13\% of the attack target web objects have sizes between 1k and 10k, 20\% between 100 and 1k, and 13\% between 10 and 100 bytes. Unlike the whole data set, the proportion of pipelined web objects to the number of web objects in the different ranges is smaller: 19\% for size range ]10,100], 12\% for ]100,1k], 15\% for ]1k,10k] and 9\% for web objects larger than 10k. The largest proportion for pipelining is in the range ]10,100], while for the whole data set this is range ]1k, 10k]. 

\subsection{Number of Web Objects}

Our attack found 92.7k web objects, 4.8k more than the number of web objects in the target TCP streams. This difference can be explained in three parcels.  1) Our attack found 1.1k web objects in 0.8k TCP streams that do not carry web objects, resulting in an additional 1.1k web objects. 2) Our attack was able to find the correct number of web objects in 28.8k TCP streams. The attack underestimated the number of web objects in 2.3k TCP streams, mostly (96\%) by only one web object less. The attack overestimated the number of web objects in a similar number of TCP streams (2.3k), however with 30.0\% of the attacks to TCP streams yielding 3 or more web objects than the actual number. The difference between the number of overestimated (6.9k) and underestimated (2.5k) web objects is an additional 4.3k web objects. 3) Finally, the attack did not find any web objects for 0.4k TCP streams that do carry web objects; this corresponds to 0.6k fewer web objects. Summing parcels 1 and 2 and subtracting parcel 3 results in the 4.8k more objects found in the attack. 

\begin{figure}[h!]
\begin{center}
\includegraphics[width=0.5 \textwidth]{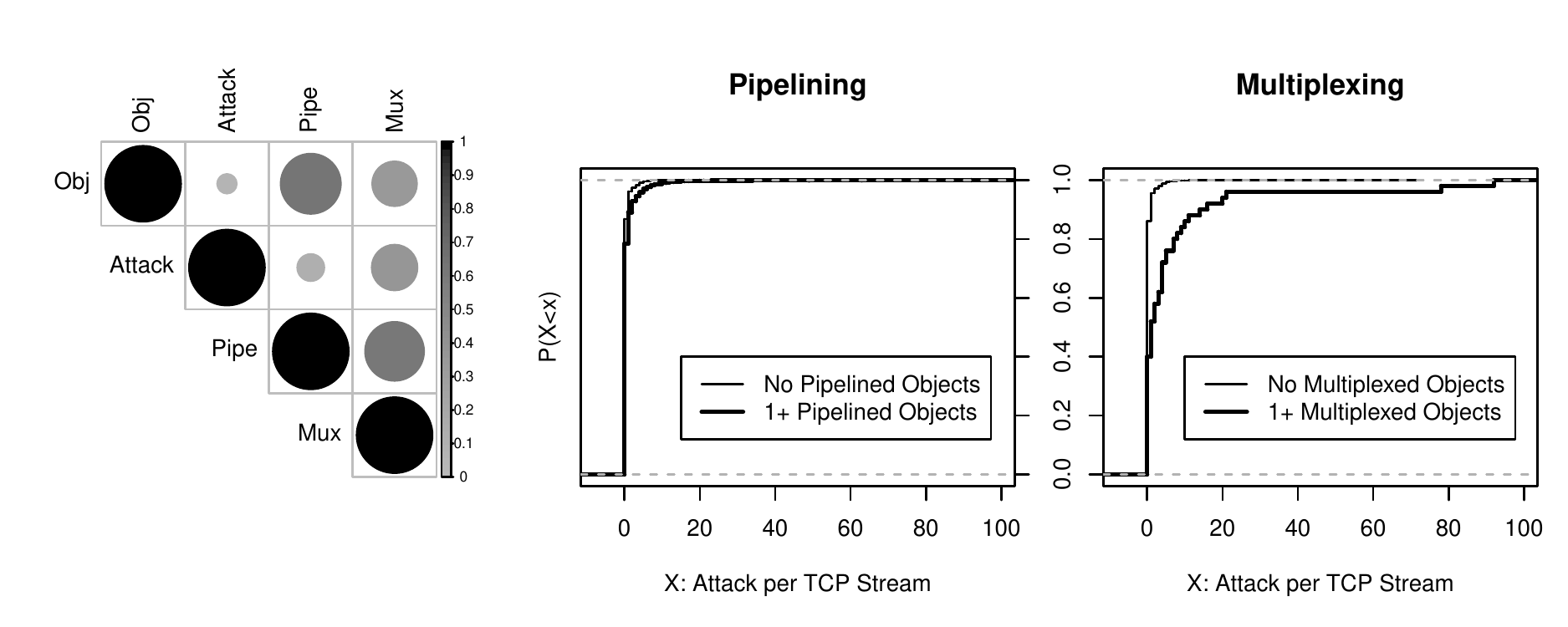} 
\end{center}
\caption{Left: Correlation between the number of web objects ($Obj$), the difference between $Obj$ and the number of web objects found in the attack ($Attack$), and the number of pipelined ($Pipe$) and multiplexed ($Mux$) web objects - all per TCP stream. Middle and right: comparison of the distribution of the $Attack$ variable per TCP streams for a) no pipelining objects , b) one or more pipelined objects, c) no multiplexed objects, d) one or more multiplexed objects.}
\label{attackobjcntdiff-pipobjcnt-muxobjcnt}
\end{figure}

We now try to understand the difference between the number of web objects in a TCP stream and the number of web objects found in the attack to that stream, which we call $Attack$. To do so, we explore the relation between $Attack$ and the number of pipelined and multiplexed objects in each TCP stream. Figure \ref{attackobjcntdiff-pipobjcnt-muxobjcnt} shows a stronger correlation between the $Attack$ variable and the number of multiplexed objects than with the number of pipelined objects. The CDFs in Figure \ref{attackobjcntdiff-pipobjcnt-muxobjcnt} confirm this; the difference between the $Attack$ distributions for TCP streams with and without multiplexing is larger than for TCP streams with and without pipelining. This makes sense since our attack is related to Assumption 2 that covers pipelining but not multiplexing.

\subsection{Web Object Size}

\begin{figure}[h!]
\begin{center}
\includegraphics[width=0.5 \textwidth]{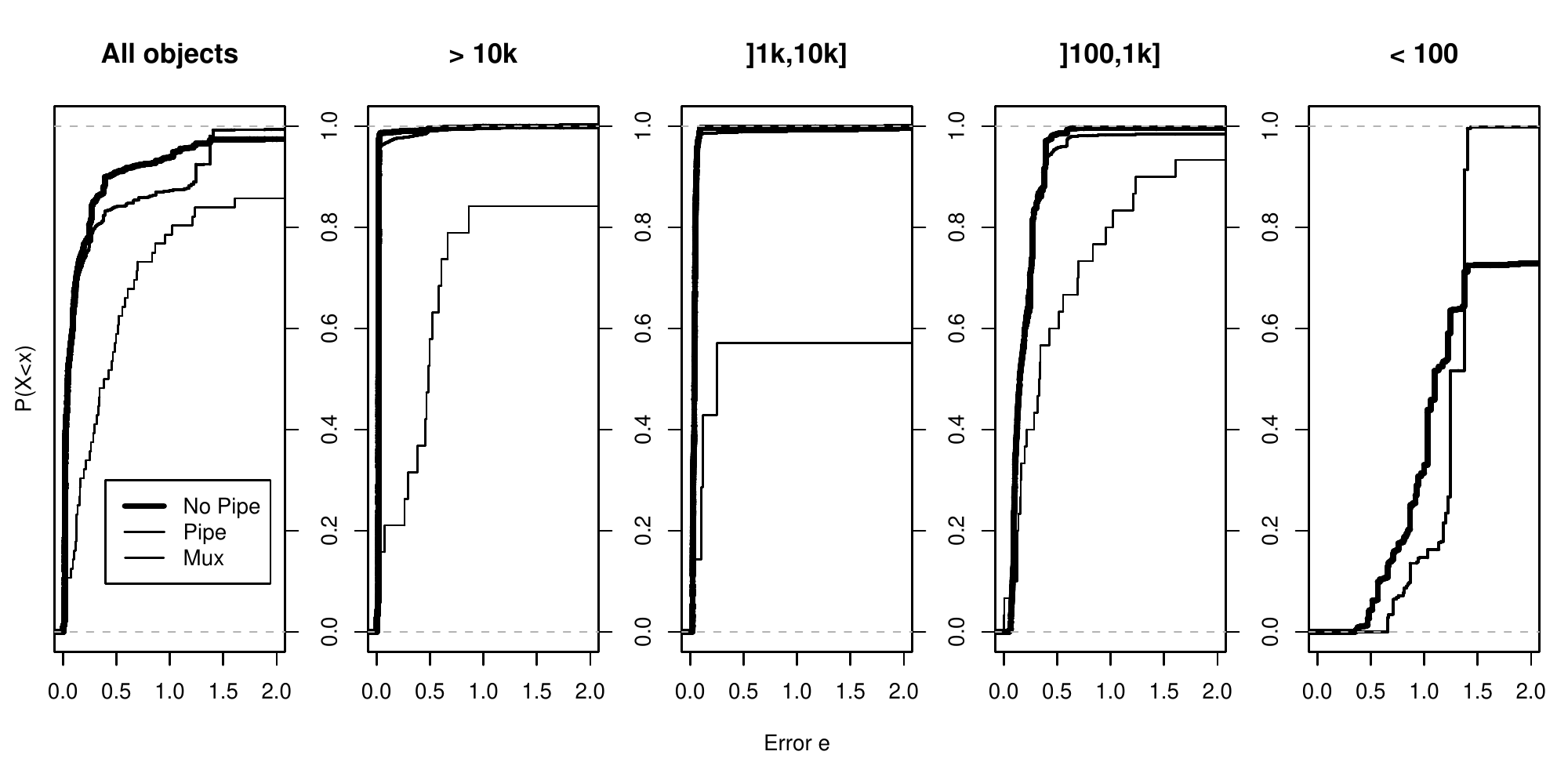} 
\end{center}
\caption{Distribution of the error $e$ of the attack (cf. section \ref{sec:error}). This distribution is shown for 1) all objects (left) and for objects with size in the ranges in bytes shown for each of the 4 images on the right and 2) objects that were not pipelined (thick line), objects that were pipelined (normal line), and obejcts that were multiplexed (thin line). Notice that no objects smaller than 100 bytes were multiplexed, thus only thick and normal lines are visible on the image on the right. }
\label{attack}
\end{figure}

We take all TCP streams for which our attack resulted in the correct number of web objects in the stream and analyze the error of the web object sizes found in the attack. To map the web object sizes found in the attack with the actual web objects in the TCP stream we order both sets of response sizes by header response timestamp and assign the first web object found in the attack to the first web object in the stream, and so forth. We then take estimated web object size and actual web object size and apply error $e$ as defined in section \ref{sec:error}. Figure \ref{attack} shows the resulting distributions of error $e$, grouped by web object size and whether the web object was pipelined, multiplexed, or neither.

We observe that non-pipelined and pipelined object error distributions are very similar. The exception are the distributions for web objects smaller than 100 bytes, which cause the difference between pipelined and non-pipelined distributions for all objects between $e =1.0$ and $e=1.5$. The error distributions for pipelined web objects are much closer to zero than the error distribution for multiplexed objects. For web object sizes larger than 1k in particular, the distributions for pipelined object error $e$ are extremely close to zero - more than 94\% of web objects larger than 1k have error $e$ smaller than 6\% and more than 95\% larger than 10k have error smaller than 3\%. Distributions for multiplexed objects show higher error than for non-pipelined and pipelined objects. This is expected given that our attack is related to assumption 2 which covers pipelining but not multiplexing. 

\section{Related Work}
\label{sec:rw}

We are not aware of any related work that tries to understand the effect of pipelining and response multiplexing in estimating HTTP/2.0 response sizes. The extent of pipelining and multiplexing on a set of web pages on the Internet is also not discussed in any work that we are aware of. In this section we go through related work in areas close to other aspects of our work -- namely side channel attacks on encrypted channels and response size estimation.

\subsection{Encrypted Channels}

Many side channel attacks have been proposed on different encrypted channels. A decade ago, and motivated by the CPU-intensive nature of application traffic classification techniques such as snort~\cite{roesch_snort_1999} and by the legal difficulties of analyzing traffic content, researchers started using features of the traffic that did not rely on its content -- namely packet size, direction, and timing~\cite{bernaille_traffic_2006}. Quickly researchers realized these features could be used together with machine learning techniques~\cite{dainotti_classification_2008} to classify encrypted traffic, which snort cannot not. Experiments for classifying SSH-tunneled application traffic~\cite{dusi_using_2009} or identifying HTTPS webmail traffic~\cite{schatzmann_digging_2010} have been motivated mostly by network management requirements, for example the need to provide different quality of service for applications with different traffic requirements, which existing methods were unable to do for encrypted traffic. 

Another and possibly more pressing motivation for side channel attacks on encrypted channels is an attack on user privacy and the possibility that some information related to application data and user behavior is leaked through the channel, which could be used against its owner. The channel that possibly raises most privacy concerns is TOR~\cite{dingledine2004tor}, which is designed precisely to protect user anonymity. Numerous attacks on TOR have been reported~\cite{sanchez2017onions, Mittal__Stealthy__2011} focusing for example on ingress-egress traffic correlation, the specifics of the TOR onioning protocol including circuits and layers, and active attacks from compromised browser or TOR nodes. The goal for attacks on TOR could be inferring the type of application being used~\cite{he_inferring_2014} or identifying the user. Attacks to more traditional VPN tunnels with the aim of identifying web pages have been described~\cite{shi_website_2014} and use traffic features such as burst size and surge period. Encrypted voice and video traffic using voice- and video-specific protocols has also been subject to side channel attacks, with the intent of uncovering spoken sentences in encrypted VoIP communications~\cite{wright_spot_2008} or identifying the video that a user last watched~\cite{dubin_i_2016}.  Finally, privacy attacks on wireless channels have also been reported, focusing e.g. on mobile application traffic classification over WLAN with WPA2 encryption~\cite{wang_i_2015} and on commercial smart home sensors with 802.15.4 wireless interfaces~\cite{copos_is_2016}. Attacks on wireless channels are especially motivating as the attacker is not required to enter private target premises and unlawfully tap the target network. 

Recently, one of the major aspects of network management -- security -- is starting to again drive side-channel attack research. As network managers face increasingly more encrypted traffic transiting their networks which they do not have access or authorization to decrypt -- either from customers or different departments in their organization, so grows the relevance of side channel attack techniques to identify intrusions over encrypted traffic. This may be especially relevant in a cloud environment where tenants deploy wide range of modern HTTP applications or in an IoT environment where HTTPS-capable yet resource-limited and possibly less frequently updated smart devices receive sensor and actuation requests from anywhere on the Internet.

HTTP over TLS is possibly the mostly used of encrypted channels. We cover HTTP over TLS next, focusing on web object response sizes and how they can be estimated by an attacker.

\subsection{Estimating Response Size and Identifying Web Sites}

A in-depth and interesting discussion on size-exposing techniques can be found in~\cite{van_goethem_request_2016}. Starting with the initial warnings that eavesdropping encrypted communications  could reveal information about its content~\cite{wagner_analysis_1996} and examples of attacks to identify web sites using fingerprinting~\cite{hintz_fingerprinting_2002}, the authors are quick to focus on online social networks. In addition to identifying which web page the user is accessing,
the state of the user could be inferred using the set of web resources that social network web sites send over the Internet. Estimating the size of web objects can now have much more severe consequences than a decade or two ago as it more directly exposes the user.

\cite{chen_side-channel_2010} illustrates the possibly high impact consequences of estimating response sizes in health-care, tax, investment, and online search web sites. Earlier, \cite{sun_statistical_2002} had shown that very good performance could be achieved in an attack to identify web sites from a set of 100k different sites, although web pages with dynamic content could be a problem. Although the issue of identifying dynamic web pages has been addressed since that~\cite{shi_website_2014}, it remains an open issue given the limited set of web pages that was studied. \cite{van_goethem_request_2016} provides a more detailed study on dynamic pages and are able to retrieve the exact size of web objects but their solution requires an active attack to be performed, which may not always be feasible. 

Pipelining is referred to in~\cite{sun_statistical_2002} as a mechanism that could prevent web object size estimation. However, at that time pipelining was not used despite having been defined in the HTTP protocol. Otherwise, related work assumes it is straightforward to obtain adequate estimations of web object sizes from the stream of TLS records.

\section{Conclusion}
\label{sec:conclusions}
The effect of pipelining and multiplexing on the estimation of web object sizes yields a perceivable increase in the relative object size estimation error. However, 1) attacking pipelining is viable as we've shown for some server IP addresses and 2) the current extent of multiplexing -- and to a lesser degree that of pipelining -- is limited. These two issues suggest that despite the potential of pipelining and multiplexing, their use needs to increase if they're to better help prevent web object size estimation.

We see many interesting related research issues ahead. The first issue is to understand to which extent it is possible to automatically detect new types of web servers and attacks, e.g. by correlating TLS record sizes with the beginning and end of HTTP responses. The second issue is to better understand why pipelining and multiplexing are used or not in apparently similar circumstances and for different captures of the same sequence of requests. We plan to develop a controlled environment where pipelining and multiplexing can be adjusted and different web applications can be installed. This will also allow us to observe the trade-off between object size estimation and quality of service. This leads to the third research issue which is to explore different countermeasures at the web engine, web application, browser, and service provider that can reduce the effectiveness of web object size estimation. Finally, the fourth research issue we see is related to measuring the effect of pipelining and multiplexing in reducing the efficiency of web page identification, for example using attack data and developing features that consider specific temporal sequences of web object sizes in each TCP stream.

\bibliographystyle{unsrt}
\bibliography{http2mux}

\end{document}